\documentclass[prl,reprint,aps,amsmath,amssymb,twocolumn,superscriptaddress,longbibliography]{revtex4-1}
\usepackage{graphicx,bm,color,appendix}
\usepackage[colorlinks,bookmarks=true,citecolor=blue,linkcolor=blue,urlcolor=blue]{hyperref}
\usepackage{times}
\usepackage{amsmath,amssymb}
\usepackage{comment}
\usepackage{appendix}

\usepackage{xcolor}
\definecolor{darkblue}{rgb}{0.,0.,0.4}
\definecolor{darkred}{rgb}{0.5,0.,0.}
\definecolor{BlueViolet}{RGB}{138,43,226}
\definecolor{SkyBlue}{RGB}{30,144,255}
\definecolor{DarkGreen}{RGB}{0,100,0}

\usepackage[normalem]{ulem}
\usepackage{comment}

\renewcommand{\vec}[1]{\bm{#1}}

\begin{document}

\title{ Conformal Operator Content of the Wilson-Fisher Transition on  Fuzzy Sphere Bilayers }

\author{Chao Han}
\author{Liangdong Hu}
\author{W. Zhu}
\affiliation{Institute of Natural Sciences, Westlake Institute for Advanced Study, Hangzhou 310024, China}
\affiliation{Department of Physics, School of Science, Westlake University, Hangzhou 310030, China }

\begin{abstract}
The Wilson-Fisher criticality provides a paradigm for a large class of  phase transitions in nature (e.g., helium, ferromagnets). In the three dimension, Wilson-Fisher critical points are not exactly solvable due to the strongly-correlated feature, so one has to resort to non-perturbative tools such as numerical simulations. Here, we design a microscopic model of Heisenberg magnet bilayer and study the underlying Wilson-Fisher $\mathrm{O}(3)$ transition through the lens of fuzzy sphere regularization. We uncover a wealth of crucial information which directly reveals the emergent conformal symmetry regarding this fixed point. In specific, we accurately calculate and analyze the energy spectra at the transition, and explicitly identify the existence of a conserved Noether current, a stress tensor and relevant primary fields.  Most importantly, the primaries and their descendants form a fingerprint conformal tower structure, pointing to an almost perfect state-operator correspondence. Furthermore, by examining the leading rank-4 symmetric tensor operator, we demonstrate the cubic perturbation is relevant, implying the critical $\mathrm{O}(3)$ model is unstable to cubic anisotropy, in agreement with the renormalization group and bootstrap calculations. The successful dissection of conformal content of the Wilson-Fisher universality class extends the horizon of the fuzzy sphere method and paves the way for exploring higher dimensional conformal field theories. 
\end{abstract}

\maketitle


\textit{Introduction.---}
Continuous phase transitions and corresponding critical phenomena exhibit remarkable universal macroscopic properties.
To understand the origin of universality, Wilson and Fisher  firstly worked out a set of fixed points in the critical $\mathrm{O}(N)$ model ($N=1,2,3,...$) \cite{WilsonFisher1972}, relevant for phase transitions in entangled polymers, helium and Heisenberg magnets \cite{Sachdev_book,Cardy_book}.
Traditionally, the derivation of these universal quantities has relied on perturbative theoretical methods \cite{WilsonFisher1972,largeN1974,Vicari2000} and brute-force numerical simulations on microscopic models \cite{Hasenbusch2023}. Remarkably, if the fixed point exhibits emergent conformal symmetry \cite{polyakov1970conformal}, the critical phenomena can be interpreted within the realm of conformal field theory (CFT) \cite{yellowbook,Belavin1984}. Such understanding is paramount for both high-energy and condensed matter physics communities.

On the other hand, the general proof  of a phase transition to be conformal invariance is extremely challenging \cite{polyakov1970conformal,polchinski1988scale,Dymarsky2015scale}, and even the evidence is very limited, especially in dimension higher than 2D (or equivalently, 1 + 1D).
A detour is to compare critical exponents from experimental measurements \cite{Vicari2002} and  Monte Carlo simulations \cite{Hasenbusch2023}
with the results obtained by numerical conformal bootstrap \cite{Rychkov:2009ij,ElShowk:2012ht,RMP_CB,vector_CB} which  explicitly assumes conformal symmetry. An additional evidence is the low-energy spectra of some discrete lattice models consistently match critical $\mathrm{O}(N)$ field theory under the $\varepsilon-$expansion \cite{Schuler2016Universal,Whitsitt2017}. 
Moreover, a more compelling  evidence is to directly expose the underlying CFT algebra and operator content. 
For instance, a celebrated feature of CFT is, for a Hamiltonian living on $S^{d-1}\times \mathbb{R}$ space-time geometry,  
the scaling dimensions of CFT operators has one-to-one correspondence with the eigen-energies of CFT states, dubbed as the state-operator correspondence \cite{Cardy1984,Cardy1985}, which is guaranteed by the conformal invariance.
Nevertheless, the simulation on curved spherical geometry $S^{d-1}$ is very challenging, despite of several attempts using stereographic projection \cite{Deng2003} and finite element discretization \cite{Brower2014Improved}. 
Very recently, this technical obstacle has been removed
by using the idea of the fuzzy sphere regularization  \cite{ZHHHH2022}. By applying this newly developed scheme to the 3D Ising critical point,  
which is equivalent to the Wilson-Fisher critical model ($N=1$) with the discrete $Z_2$ symmetry, the conformal data including the scaling dimensions \cite{ZHHHH2022} and operator product expansion (OPE) coefficients \cite{Hu2023} have been unambiguously characterized. With this exciting progress, it is highly desired to apply the fuzzy sphere microscope to the general $\mathrm{O}(N)$ model ($N>1$) with the continuous symmetry, where the conformality has rarely been explored before. 

In this paper, we construct a (2+1)-D model on the quantum fuzzy sphere and 
we show that the model realizes a continuous quantum phase transition belong to three-dimensional classical Heisenberg universality class.
The location of critical point is accurately determined by the finite-size scaling of magnetic order parameter assisted by the data collapse and crossing-point analysis.
Remarkably, clear signatures of conformal invariance are observed at the critical point, through uncovering the emergent state-operator correspondence. 
We identify the conserved Noether current, stress-energy tensor and relevant conformal primaries and their descendants  in the operator spectrum. 
The conformal data including the scaling dimensions and the OPE coefficients are crucial for understanding of the instability of this Wilson-Fisher $\mathrm O$(3) fixed point, e.g. 
we demonstrate that the lowest rank-4 symmetric operator is more likely relevant, pointing to relevance of cubic perturbation. 
Additionally, we elucidate that the proposed fuzzy sphere model can be feasibly generalized to study other $\mathrm O$(N) classes (e.g. three-dimensional XY transition).

\begin{figure}[!b]
	\includegraphics[width=0.3\textwidth]{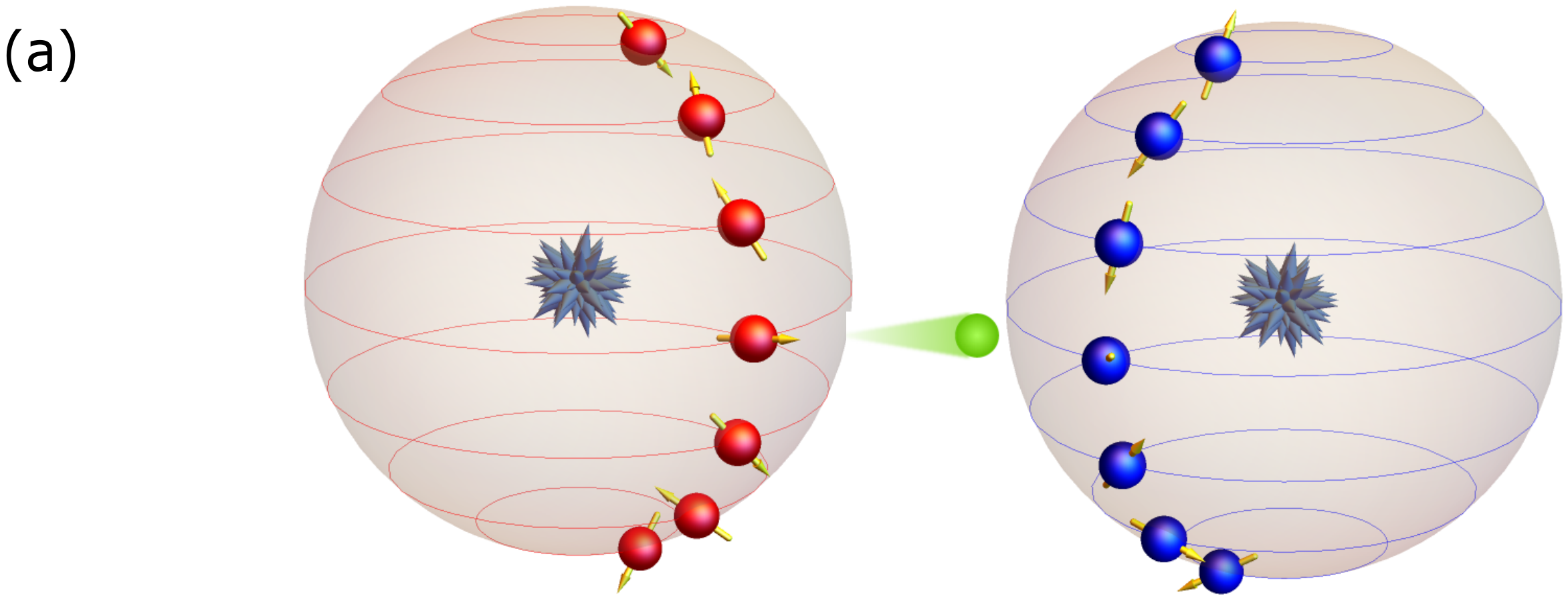}
	\includegraphics[width=0.45\textwidth]{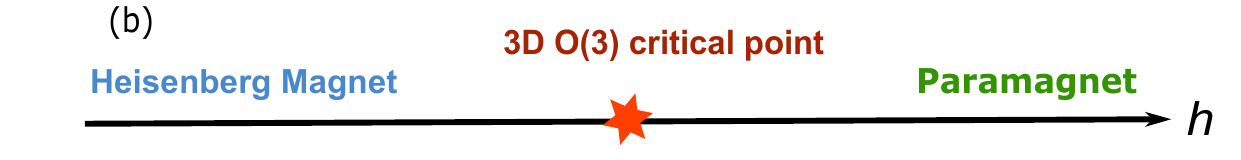}
	\includegraphics[width=0.225\textwidth]{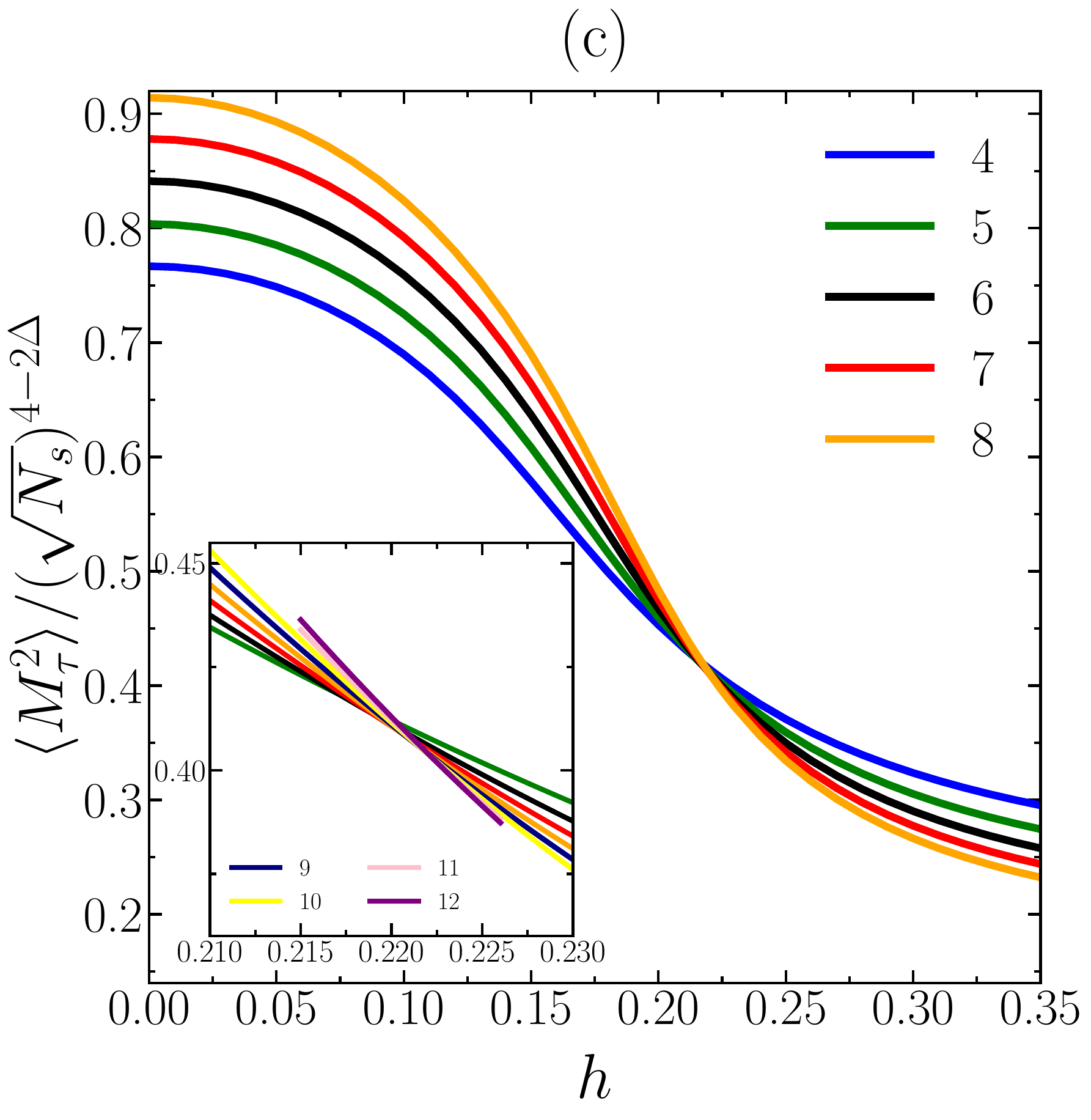}
	\includegraphics[width=0.225\textwidth]{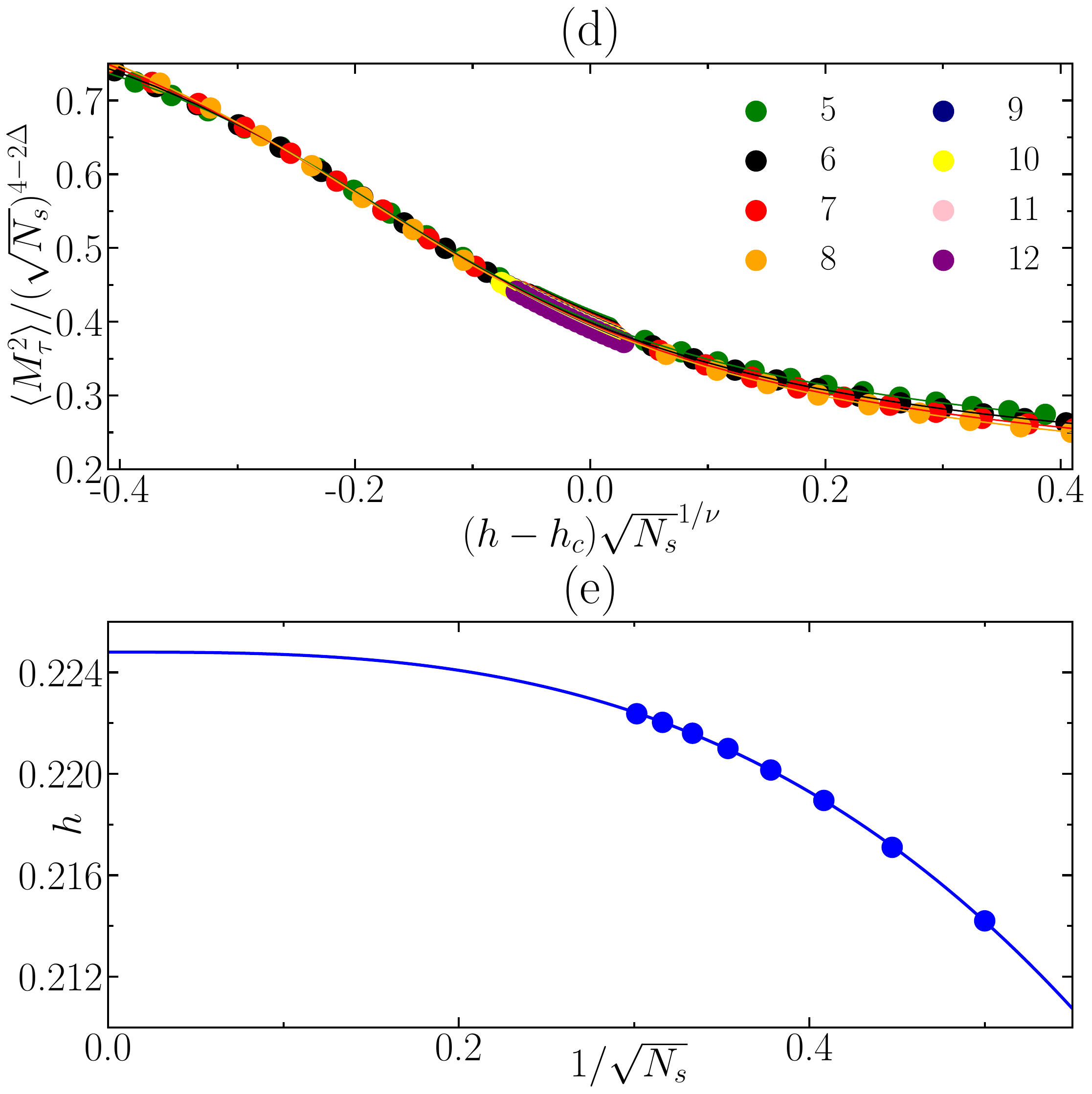}
	\caption{ (a) Sketches of the bilayer fuzzy sphere model:  Interacting fermions move on a fuzzy sphere bilayer; and the fermion is able to tunneling between two layers. 
		(b) A schematic plot of phase diagram with a critical point separating a paramagnet from a symmetry breaking Heisenberg magnet. 
		(c) Finite size scaling of order parameter $\langle M^2_{1(2)} \rangle/N_s^{2-\Delta_\phi}$, where $\Delta_\phi=0.519$ is the scaling dimension of the $\mathrm O(3)$ vector field relating to the critical exponent $\eta= 2\Delta_\phi-1$. $N_s=2s+1$ is the number of Landau orbitals (i.e. Heisenberg spins), which relates to the length scale radius as $R\sim\sqrt N_s$.  
		(d) The data collapse of the rescaled order parameter
		according to $ f((h-h_c) L^{1/\nu})$ with $\nu=1/(3-\Delta_s)$ and $\Delta_s \approx 1.595$, where $h_c$ is a free fitting parameter. The best fit gives $h_c \approx 0.225$. (e) Finite size scaling of crossing points by a finite-size pair $(N_s,N_s+1)$ gives rise to an extrapolated value $h_c \approx 0.2248 \pm 0.0001$. 
  }
	\label{fig:phasediagram}
\end{figure}

\textit{$\mathrm{O}(3)$ transition on the fuzzy sphere.---}
The fuzzy sphere regularization~\cite{ZHHHH2022}  describes interacting fermions moving on a sphere with a $4\pi s$ magnetic monopole at the origin \cite{Sphere_LL_Haldane}. Owing to the monopole, the kinetic energy of fermions forms quantized Landau levels and each orbital is described by the monopole Harmonics $Y^{(s)}_{n+s, m}(\vec\Omega)$~\cite{WuYangmonopole} ($n=0, 1, \cdots$ denotes the Landau level index, and $\vec\Omega=(\theta, \varphi)$ are the spherical coordinates). By tuning the interactions, the spin degree of freedom of fermions would undergo a phase transition.
To apply the fuzzy sphere to a critical $\mathrm{O}(3)$ phase transition, we borrow the idea from the prototypical spin-$1/2$ Heisenberg bilayer model consisting of intralayer and interlayer interactions \cite{Sandvik1994,LWang2006,doublelayerQHE_Sarma,sm}. 
We introduce four-flavor fermions $\mathbf{\Psi}=(\psi_{1\uparrow}, \psi_{1\downarrow},\psi_{2\uparrow},\psi_{2\downarrow})^T$ with layer $\tau=1,2$ and spin $\sigma=\uparrow,\downarrow$ indices living on the fuzzy sphere (see Fig. \ref{fig:phasediagram}(a)).
Consequently, we consider a real-space Hamiltonian:
\begin{widetext}	
\begin{align*}
	H_\textrm{int}=\! \int\!d\vec{\Omega}_{a,b}\! \left[U_0 n(\vec{\Omega}_a) n(\vec{\Omega}_b)+ U_2 \vec{n}_1(\vec{\Omega}_a) \cdot \vec{n}_2(\vec{\Omega}_b) 
	- U_1 (\vec{n}_1(\vec{\Omega}_a) \cdot \vec{n}_1(\vec{\Omega}_b) + \vec{n}_2(\vec{\Omega}_a) \cdot \vec{n}_2(\vec{\Omega}_b)) 
	  \right]  \!
	- h \int d\vec{\Omega} \hat{\mathbf{\Psi}}^\dagger \tau^x\sigma^0 \hat{\mathbf{\Psi}},
\end{align*}
where the local density operator of layer-$\tau$ is $\vec{n}_\tau(\vec \Omega) = (n_\tau^x,n_\tau^{y},n_\tau^z) =\psi^\dagger_\tau(\vec \Omega) \vec{\sigma} \psi_\tau(\vec \Omega)$ and the total density is $n(\vec \Omega)=\mathbf{\Psi}^\dagger(\vec \Omega)\mathbf{\Psi}(\vec \Omega)$. For simplicity we consider the potentials to be short-ranged interactions $U_0=\delta(\vec{\Omega}_1-\vec{\Omega}_2)$, $U_1=u_1\delta(\vec{\Omega}_1-\vec{\Omega}_2)$ and 
$U_2=u_2 \nabla^2 \delta( \vec{\Omega}_1-\vec{\Omega}_2 )$, and we set $u_1=0.55,u_2=0.19$, which are optimized to minimize the finite-size effect. The transverse field strength $h$ controls tunneling effect between two different layers.

\end{widetext}

In practice, we consider the second quantization form of this model by the projecting $H_\textrm{int}$ to the lowest Landau level (see Supple. Mat. Sec. A \cite{sm}), 
using $\psi_{\tau\sigma} (\vec{\Omega}) = \frac{1}{\sqrt{N_s}} \sum_{m=-s}^s \hat{c}_{m, \tau,\sigma}  Y^{(s)}_{s, m} (\vec{\Omega})$. 
Here the number of Landau orbitals $N_s=2s+1$ plays the role of system size $N_s\sim R^2$ ($R$ is the radius of sphere). We consider the Landau level is filled by $2 N_s$ electrons in total.
This model possesses the $\mathrm{SO}(3)$ rotation symmetry of sphere,  the global $\mathrm{O}(3)$ symmetry, the layer inversion symmetry $\mathbf{I}_v: \hat c_{m,\alpha,\sigma}\to \tau^x_{\alpha\beta}\hat c_{m,\beta,\sigma}$, and the particle-hole symmetry $\mathbf{P}: \hat{c}_{m,\alpha,\sigma}\to i\tau^y_{\alpha\beta} \hat{c}^*_{m,\beta,\sigma}, i\to-i$, in addition to the $U(1)$ charge conservation. 
Numerically, this model is solved using exact diagonalization and density matrix renormalization group (DMRG) \cite{SWhite1992,Feiguin2008}. We perform DMRG calculations with bond dimensions up to $D=6000$, and we explicitly impose three $U(1)$ symmetries, i.e., spin-resolved fermion number and $z$-component angular momentum. 
For the largest system size $N_s=12$, the maximum truncation errors for the ground state is less than $3.68 \times 10^{-7}$.

The phase transition is obtained by the conventional finite-size scaling of the order parameter (i.e. $\mathrm{O}(3)$ vector), which inspects the spontaneous symmetry breaking 
\begin{equation}
	\vec{M}_\tau=\sum_{m\alpha\beta} \hat{c}^\dagger_{m,\tau,\alpha} \vec{\sigma}_{\alpha\beta} \hat{c}_{m,\tau,\beta}.
\end{equation}
Utilizing this order parameter, our calculation confirms a direct continuous phase transition from ordered Heisenberg magnet to disordered paramagnet in the proposed model. 
At the transition point, the order parameter should scale as $\langle M^2_\tau \rangle \sim R^{4-2\Delta_\phi} = N_s^{2- \Delta_\phi}$ \cite{Hasenbusch2023,Sandvik1994,LWang2006,Ippoliti2018Half},
where $\Delta_\phi\approx 0.51928$ is the scaling dimension of $\mathrm{O}(3)$ vector field \cite{RMP_CB}. 
Fig.~\ref{fig:phasediagram} (c) illustrates the rescaled order parameter $\langle M^2_\tau \rangle/N_s^{2-\Delta_\phi}$ with respect to the transverse field strength $h$ for various $N_s$. 
$\langle M^2_\tau \rangle/N_s^{2-\Delta_\phi}$ is almost unchanged near the crossing point $h \approx h_c$, which signals the phase
transition point.
The exact value of critical point $h_c$ can be obtained by two ways: 1) $h_c$ is the best fitting parameter for the data collapse as shown in Fig. \ref{fig:phasediagram}(d), or 2) through the crossing-point analysis, crossing points for different sizes tend to a extrapolated value $h_c$, as shown in Fig.\ref{fig:phasediagram}(e). 
Importantly, both analysis give a consistent estimation of $h_c\approx 0.225$, which is taken to be the critical point for the following discussion.   
Additionally, we have also computed the binder cumulant and the lowest energy gap, finite-size scaling of which confirm the estimation of $h_c$ (see Supple. Mat. Sec. B \cite{sm}).  
In a word, the critical behavior of order parameter shows 
the critical point in this bilayer fuzzy sphere model described by the $\mathrm O (3)$ universality class.

\begin{figure}[t]
	\centering
	\includegraphics[width=0.23\textwidth]{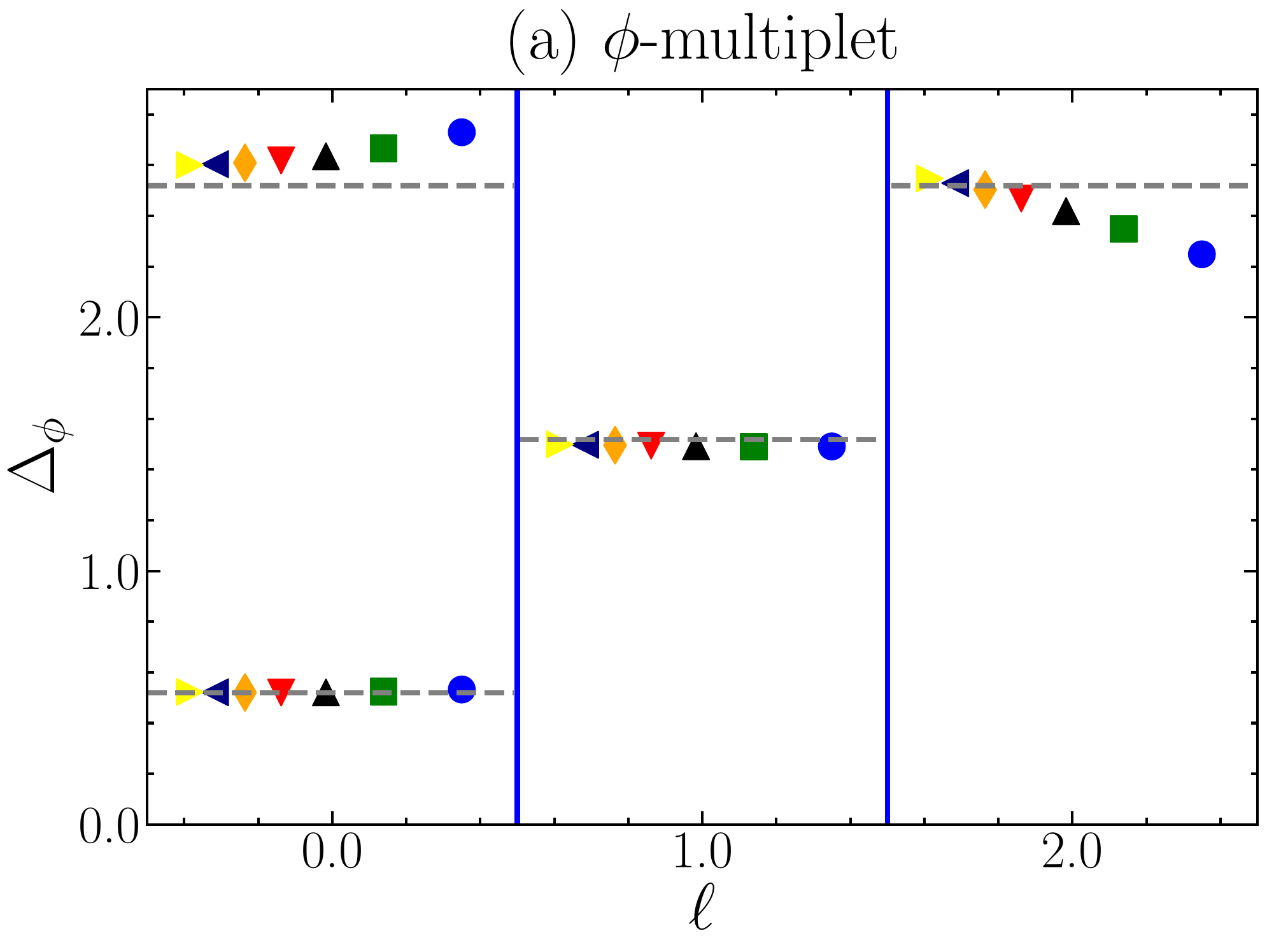}
	\includegraphics[width=0.23\textwidth]{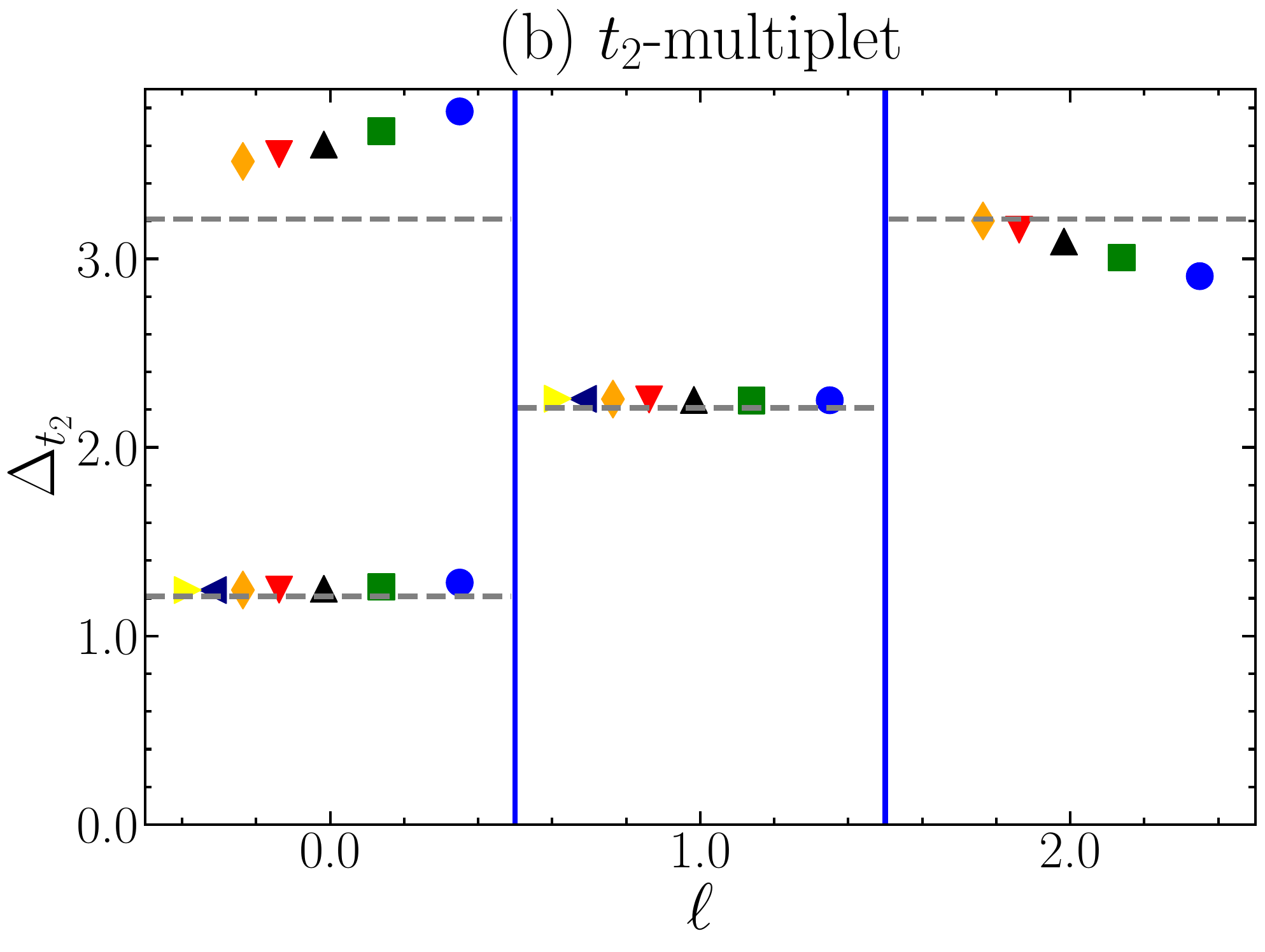}
	\includegraphics[width=0.23\textwidth]{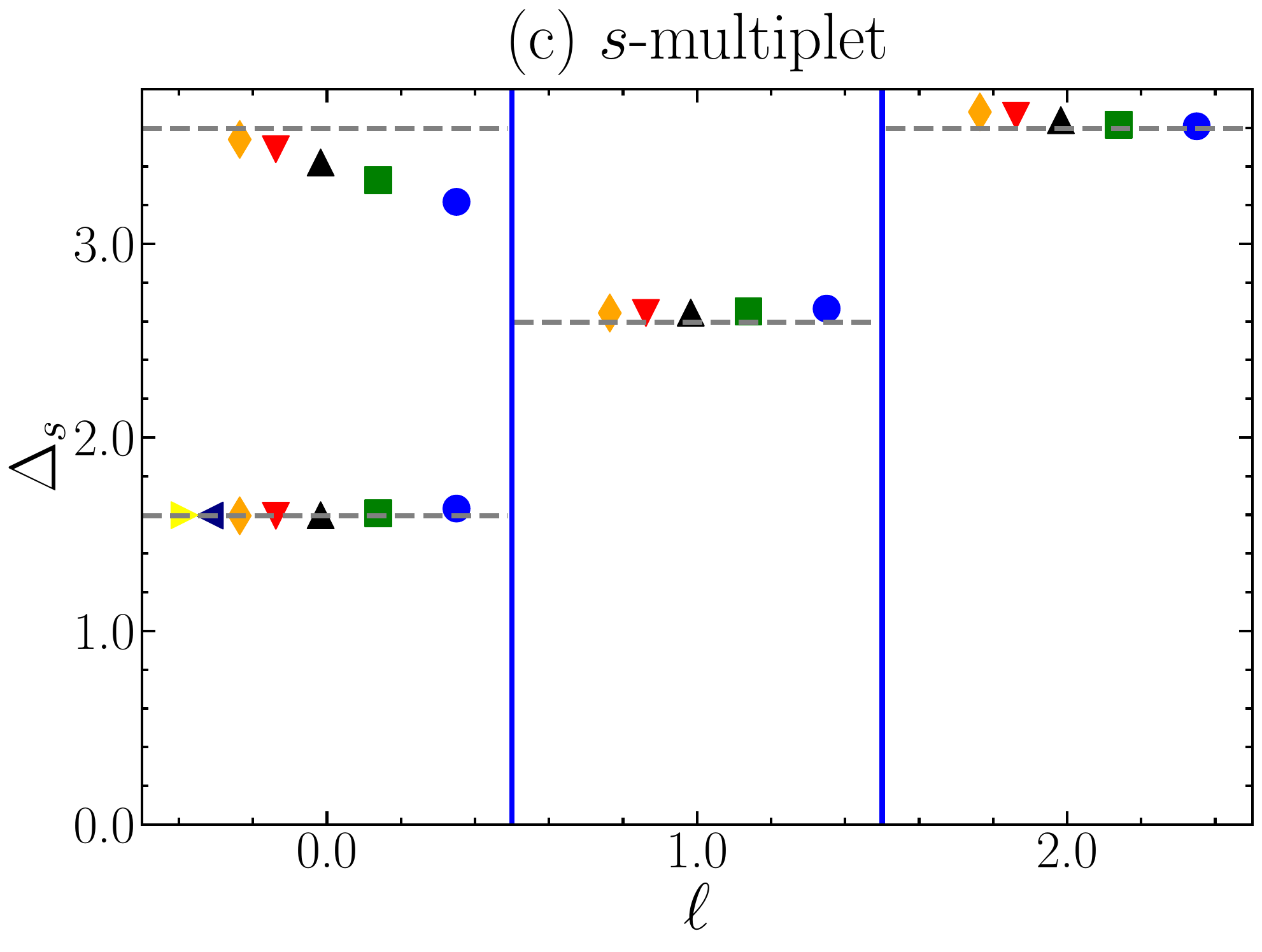}
	\includegraphics[width=0.23\textwidth]{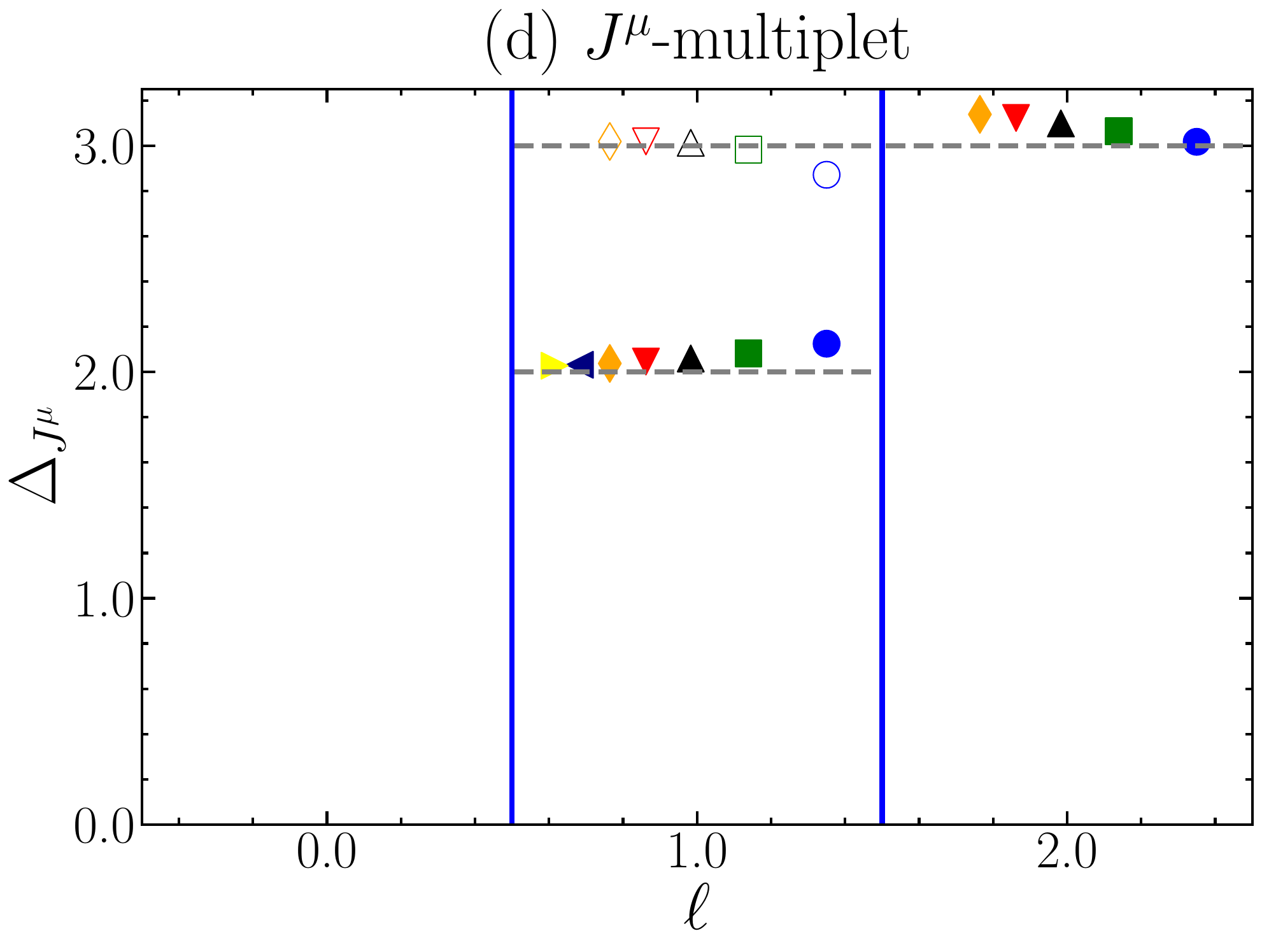}
	\caption{Conformal multiplet of several low-lying primary operators. Scaling dimension $\Delta$ versus Lorentz spin $\ell$ for (a) the lowest  vector $\phi$, (b) the lowest rank-2 symmetric traceless tensor $t_2$, (c) the lowest scalar field $s$ and (d) conserved  current $J_\mu$.  
		The plots are calibrated by the scaling dimension of the energy-momentum tensor $\Delta_{T_{\mu\nu}}=3$. 
  Solid (open) symbols represent
parity even (odd) operators. 
The dashed horizontal lines are the prediction from conformal bootstrap, and the discrepancy is relatively smaller for the primaries and their first descendants.
	}
	\label{fig:multiplet}
\end{figure}

\textit{Operator spectrum.---}
We now turn to the main results of this paper. The great advantage to work on the spherical geometry $S^2$ is, we can unlock the so-called state-operator correspondence \cite{Cardy1984,Cardy1985}, i.e. the eigen-energy gap takes the form $\delta E_n= E_n-E_0 = v \Delta_n/R$,
where $v$ is the model-dependent speed of light, and $\Delta_n$ are the scaling dimensions of the CFT operators. Therefore, we compute energy spectra at the critical point and compare it with CFT predictions.

To examine if the eigenstates in this proposed model form representations of 3D conformal symmetry, we analyze the low-lying spectra according to the following rules:
1) We rescale the full spectrum by setting the energy-momentum tensor $T_{\mu\nu}$ to be exactly $\Delta_{T_{\mu\nu}} = 3$;  2) The lowest-lying energy states that cannot be generated from other fields are identified as primaries; 3) The descendants are produced by applying raising ladder operators to the identified primary states and by matching the quantum numbers. 
For example, for a scalar primary $O$ with quantum number $(\ell,S)$ (Here $\ell$ is the quantum number of SO$(3)$ rotation symmetry of fuzzy sphere behaving as the Lorentz spin of the conformal group and $S$ is the quantum number of global spin rotation symmetry.), its descendants can be written as 
$	\partial_{\nu_1} \cdots \partial_{\nu_j} \square^n O$, which takes the scaling dimension $\Delta_O + 2n +j$ with quantum number $(\ell+j, S)$.
Fig. \ref{fig:multiplet} depicts numerically identified conformal multiplet (i.e. primary and its descendants) of the lowest $\mathrm{O}(3)$ vector $\phi$, the lowest $\mathrm{O}(3)$ traceless tensor $t_2$, the lowest $\mathrm{O}(3)$ scalar $s$ and the Noether current $J_\mu$. 
Remarkably, we found that the low-lying eigen-states approximately form representations of the 3D conformal symmetry up to small finite-size corrections.
This is a direct and unambiguous demonstration of the emergent conformal symmetry of the 3D Wilson-Fisher $\mathrm{O}(3)$ transition. To our best knowledge, such direct evidence has not been reported before. Additionally,
the numerical discrepancy is typically more significant for the fields with larger $\Delta$ and larger angular momentum $\ell$, which is attributed to the finite-size effect: The CFT only captures the low-energy effective theory for our model, and to access large angular momentum fields requires the simulation on sphere with large enough radius.

\begin{figure}[t]
	\centering
    \includegraphics[width=0.45\textwidth]{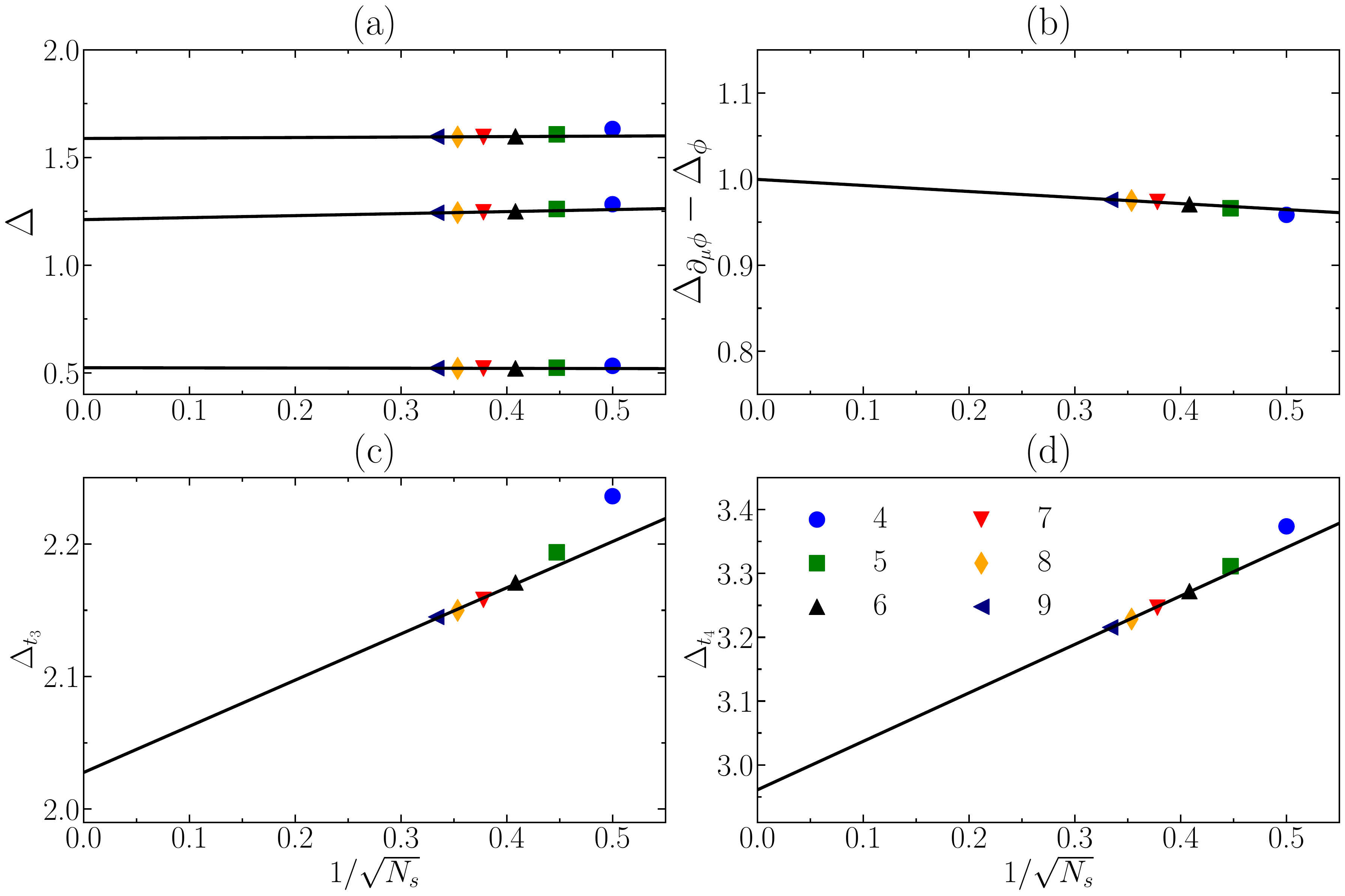}
	\caption{	
        Finite-size extrapolations of the scaling dimensions of (a) the lowest vector $\phi$, the lowest rank-2 tensor $t_2$, and the lowest scalar field $s$, (b) gap of descendant field $\Delta_{\partial_\mu\phi} - \Delta_{\phi}$, (c) the lowest rank-3 tensor $t_3$ and (d) the lowest rank-4 tensor $t_4$. Finite-size extrapolations with system sizes $N_s=6-9$ give rise to $\Delta_\phi \approx 0.524(4)$, $\Delta_{t_2} \approx 1.211(8)$, $\Delta_{s} \approx 1.588(9)$, $\Delta_{t_3} \approx 2.028(11)$, $\Delta_{t_4}\approx 2.961(12)$, and $\Delta_{\partial_\mu\phi}-\Delta_\phi\approx 1.000(1)$.
	}
	\label{fig:scaling}
\end{figure}

\begin{table}[!b]
	\setlength{\tabcolsep}{0.2cm}
	\renewcommand{\arraystretch}{1.4}
	\centering
	\caption{Low-lying primary operators identified via state-operator correspondence on the fuzzy sphere. We only take the first three digits from the data in literature. Error analysis see Sec. E. \cite{sm}. 
 } \label{tab:scaling_dims}
	\begin{tabular}{cccccc} \hline\hline
		& $\phi$ & $t_2$  & $s$ & $t_3$ & $t_4$ \\
		\hline
		$\epsilon-$exp\cite{HENRIKSSON2023,Vicari2000} & 0.510 & 1.232 & 1.610 & - & 2.911  \\
		large-$N$\cite{HENRIKSSON2023} & 0.499 & 1.339 & 1.301 & - & 3.447   \\
		Bootstrap\cite{Su2021}  & 0.519 & 1.209 & 1.595  & 2.039  & $<2.991$  \\				
		MC\cite{Hasenbusch2023} & 0.519 & 1.210 & 1.594 & 2.039 & 2.986   \\
		Fuzzy sphere & 0.524 & 1.211 & 1.588 & 2.028 & 2.961 \\			
		\hline\hline
		
	\end{tabular} 
\end{table}

After verifying the emergent conformal symmetry, we further compare scaling dimensions of the identified primary operators with the existing data from various methods \cite{Hasenbusch2023,Vicari2002,Su2021,RMP_CB,vector_CB,LWang2006,Kleinert1995,ZinnJustin1998,Vicari2000}. 
As listed  the relevant primary operators that we have identified in Tab. \ref{tab:scaling_dims}, overall we find a reasonable agreement with numerical bootstrap \cite{RMP_CB,Su2021} and Monte Carlo data \cite{Hasenbusch2023}, e.g. the averaged discrepancy from the bootstrap data is less than $1\%$. 
Despite of the small discrepancy, the precision is still sufficiently high to further increase the confidence that the universality class of the transition falls into the 3D Wilson-Fisher $\mathrm{O}(3)$ type.
In particular, these data are crucial to understand the physics of $\mathrm O(3)$ critical point. For instance, the lowest rank-4 symmetric tensor operator $t_4$ corresponds to the anisotropic cubic perturbation. This operator is dangerously relevant, according to the existing numerical computation ~\cite{Su2021,HENRIKSSON2023}. Although the finite-size value of its scaling dimension flows (Fig. \ref{fig:scaling} (d)), our calculation confirms its relevance $\Delta_{t_4} \approx 2.961(12)$.

\begin{figure}[t]
	\centering
	\includegraphics[width=0.23\textwidth]{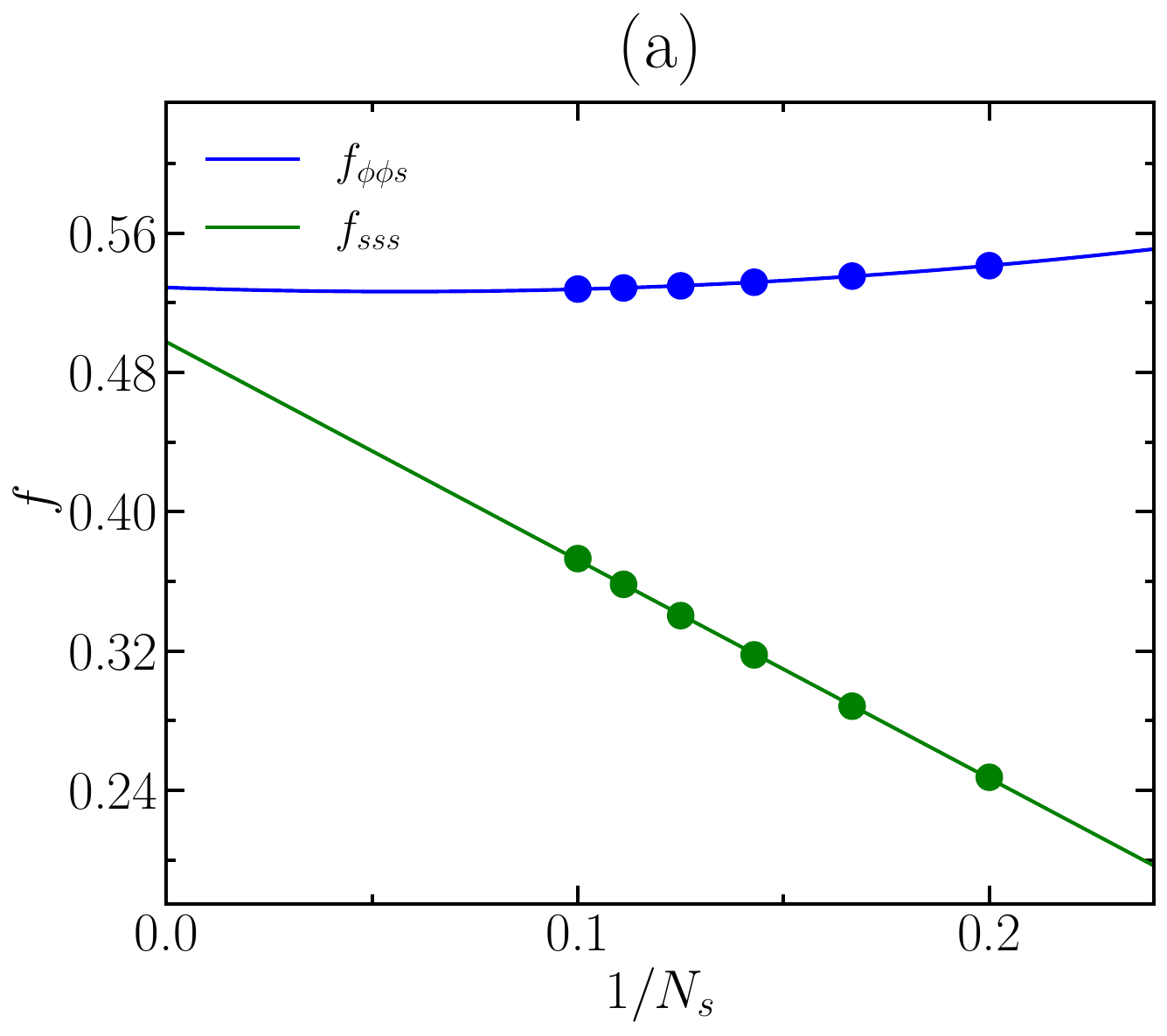}
	\includegraphics[width=0.23\textwidth]{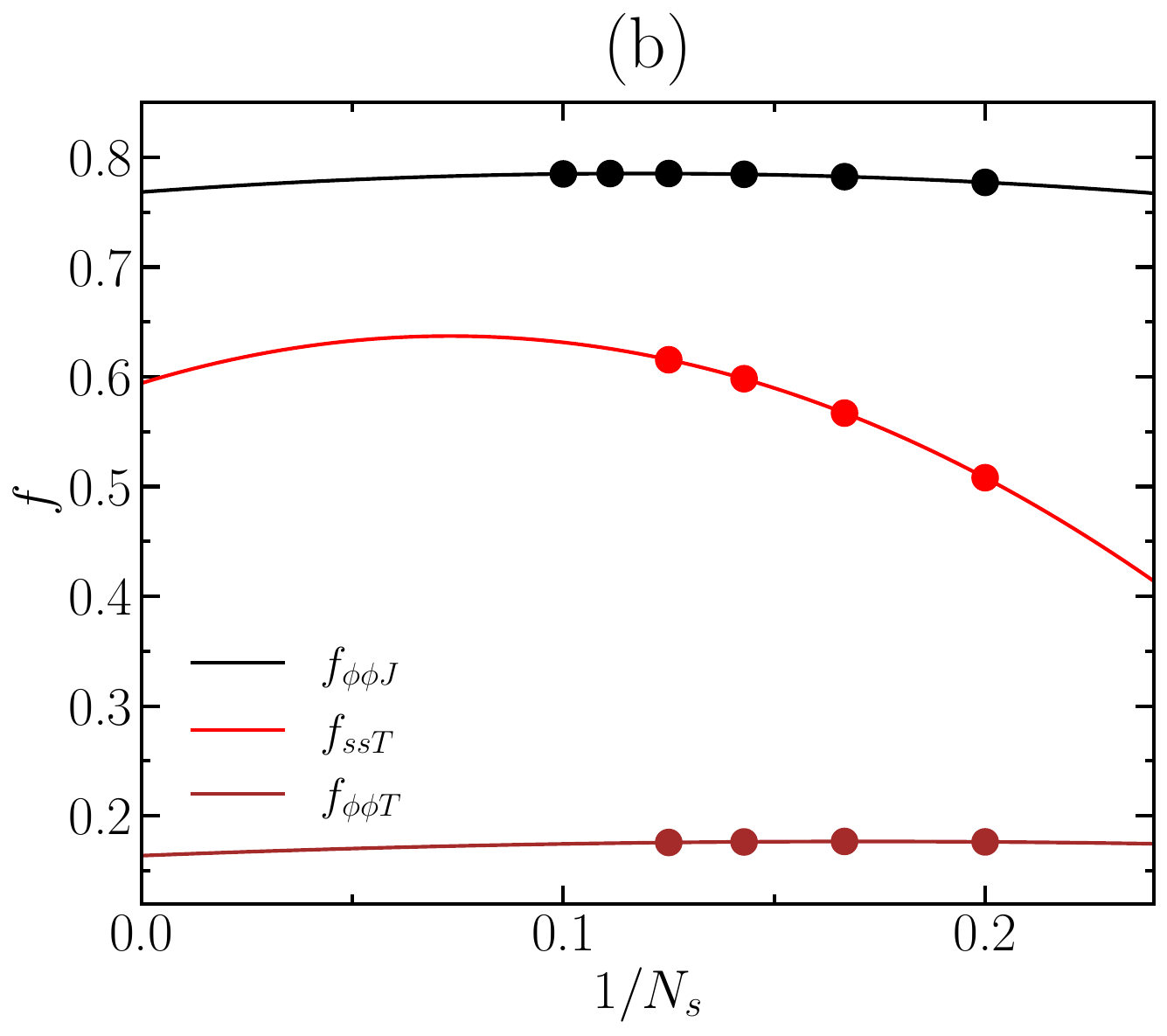}
	\caption{	
		(a) Two representative OPE coefficients involving the scalar primary $f_{\phi\phi s}$ (blue) and $f_{sss}$ (green). 
		(b) Three representative OPE coefficients involving the energy-momentum tensor $f_{\phi\phi T}$ (brown), $f_{ssT}$ (red) and the conserved current $f_{\phi\phi J}$ (black). 
        The finite-size extrapolation is up to $R^{-4}\sim N_s^{-2}$, where $N_s=5-8$ data are from the ED and $N_s=9-10$ data are from DMRG. 
	}
	\label{fig:ope}
\end{figure}

\begin{table}[b]
	\setlength{\tabcolsep}{0.2cm}
	\renewcommand{\arraystretch}{1.4}
	\centering
	\caption{List of the OPE coefficients from the fuzzy sphere and comparison with the bootstrap data \cite{RMP_CB,Su2021}. Error analysis see Sec. E. \cite{sm}.
 } \label{tab:ope}
	\begin{tabular}{cccccc} \hline\hline
		& $f_{\phi\phi s}$ & $f_{sss}$  & $f_{\phi\phi T}$ & $f_{\phi\phi J}$ & $f_{ssT}$ \\
		\hline
		Bootstrap\cite{Su2021}  & 0.524 & 0.506 & 0.189  & 0.743 & 0.580   \\				
		Fuzzy sphere & 0.525 & 0.507 & 0.169 & 0.752 & 0.578 \\			
		\hline\hline
		
	\end{tabular} 
\end{table}

\textit{Correlation functions and OPE coefficients.---}
Besides the state perspective, we can also study the operator perspective of the identified primaries. 
Here let us take the OPE coefficient $f_{\phi\phi s}$ as an example.
Since $\phi$ relates to the local operator $\vec{n}_\tau(\bm \Omega)$,   we have \cite{Hu2023}
\begin{align}
	\langle \phi | \vec{n}_\tau(\bm \Omega) |0\rangle  &= \frac{1}{R^{\Delta_\phi}}\left(c_\phi + \sum_{n=1}^{\infty} \frac{a_n}{R^{2 n}} \right), \\
	\langle \phi | \vec{n}_\tau(\bm \Omega) |s\rangle &= \frac{f_{\phi \phi s}}{R^{\Delta_\phi}}\left( c_\phi +  \sum_{n=1}^{\infty} \frac{\tilde a_n}{R^{2 n}} \right) + ... 
\end{align}
where '$...$' denotes the contribution from other primaries and associated descendants.
Thus, we can compute $f_{\phi\phi s}$ using 
\begin{equation}\label{eq:ope_phiphis}
	\frac{\langle \phi | \vec{n}_\tau(\bm \Omega) |s \rangle}{ \langle \phi |  \vec{n}_\tau(\bm \Omega) |0\rangle} = f_{\phi\phi s} + \frac{\tilde a_1 - a_1}{c_\phi R^2} + O(R^{-4}).
\end{equation}
To eliminate the two leading corrections from the descendant fields $\square \phi,\square^2 \phi$, we perform a finite-size extrapolation up to $R^{-4}$ (i.e. $N_s^{-2}$). Similarly, other OPE coefficients can be extracted within the same method \cite{sm}. 

To quantify the data of OPE, we compare our results with the numerical bootstrap data  \cite{Su2021,RMP_CB} in Tab. \ref{tab:ope}. 
The data are in overall agreement. 
The ability to access both scaling dimensions and OPE coefficients together shows the superiority of fuzzy sphere scheme.

\textit{Summary and discussion.---}
We have constructed a microscopic model of  Heisenberg magnet bilayer living on the space-time geometry $S^2\times \mathbb{R}$, which realizes a order-disorder transition belong to Wilson-Fisher $\mathrm{O}(3)$ universality class. 
At the phase transition point, clear evidence of emergent conformal symmetry is observed by identifying the one-to-one correspondence between CFT operators and the eigenstates, through which we are able to identify conformal data including the scaling dimensions and OPE coefficients of relevant primary operators. These information are crucial to understand the stability of Wilson-Fisher fixed point under various perturbations.
In addition, the proposed bilayer fuzzy sphere model can also capture the essential physics of three-dimensional XY transition belong to Wilson-Fisher $\mathrm{O}(2)$ class, by properly adjusting the global symmetry of quantum spins (see Supple. Mat. \cite{sm}). 
In this context, the current work extends the horizon of the fuzzy sphere scheme to general Wilson-Fisher universality class with continuous global symmetry, which demonstrates the vibrancy of this newly proposed method. We envision the fuzzy sphere will be a powerful tool to explore 3D CFTs in more exotic criticalities.

\begin{acknowledgments}
\textit{Acknowledgment.---}	We thank Yin-Chen He for simulation discussion and collaboration on related projects. We thank Ning Su for useful discussions. This work was supported by National Science Foundation of China under No. 92165102, 11974288. 

\textit{Note added.---} We become aware of another work on XY transition using fuzzy sphere scheme: arXiv.2310.xxxxx.
\end{acknowledgments}

\bibliographystyle{apsrev}
\bibliography{ref}

\clearpage
\appendix
\begin{widetext}

\begin{center}
\textbf{Supplementary materials}    
\end{center}

\setcounter{subsection}{0}
\setcounter{equation}{0}
\setcounter{figure}{0}
\renewcommand{\theequation}{S\arabic{equation}}
\renewcommand{\thefigure}{S\arabic{figure}}
\renewcommand{\thetable}{S\arabic{table}}
\setcounter{table}{0}

\appendix

In this supplementary material, we will show more details to support the discussion in the main text. In Sec. A, we discuss the spherical Landau levels, then derive the second-quantization form of the real-space Hamiltonian and connection to Haldane pseudopotential. 
We also explain the intuition of fuzzy sphere bilayer model for general Wilson-Fisher $\mathrm O$(N) transitions.
In Sec. B, we provide more detailed analysis on the finite-size scaling of physical observables, the binder cumulant $U_4$ and the lowest energy gap, across the phase transition. We also calculate the charge gap to demonstrate that the degree of freedom in the low-energy region is Heisenberg spin. In Sec. C,
we show how to get the OPE coefficients using spin operators and derive the finite-size scaling forms. Meanwhile, we calculate and show the result of two-point correlation function of $\phi$ in Sec. D. In Sec. E,  we present the detail of numerical data and error analysis.

\subsection{A. 1. Spherical Landau levels}
We consider a sphere with radial $R$, and a $4\pi s (2 s \in \mathbb{Z})$ monopole located in the center of the sphere. The Hamiltonian for an electron moving on the surface has only the kinetic term \cite{Sphere_LL_Haldane}
\begin{equation}
H_0=\frac{1}{2 M_e R^2} \Lambda_\mu^2,
\end{equation}
where $M_e$ is the electron’s mass and $\Lambda_\mu=\partial_\mu+i A_\mu$ is the covariant angular momentum, $A_\mu$ is the gauge field of the monopole. Here we have set $\hbar=e=c=1$. Corresponding eigen energies are known as the famous Landau levels, $E_n=[n(n+1)+(2 n+1) s] /\left(2 M_e r^2\right)$, with $n=0, 1, 2, \cdots$ the Landau level index. The $(n+1)_{\text{th}}$ Landau level has $2s+2n+1$-fold degenerate, and the single particle states in each Landau level are called Landau orbitals. 
The wave functions for each Landau orbital on LLL are called monopole harmonics~\cite{WuYangmonopole}
\begin{equation}
Y^{(s)}_{s,m}(\vec{\Omega})=N_m e^{i m \varphi} \cos ^{s+m}\left(\frac{\theta}{2}\right) \sin ^{s-m}\left(\frac{\theta}{2}\right).
\end{equation}
Here $m=-s, -s+1, \cdots, s-1, s$, the normalization factor $N_m=\sqrt{(2 s+1) ! / 4 \pi(s+m) !(s-m) !}$, and $\boldsymbol{\Omega}=(\theta, \varphi)$ is the spherical coordinates.

\subsection{A. 2. Second-quantization Hamiltonian}
Here we derive the second-quantization form of the real-space Hamiltonian. Firstly, we define the total density operator as
\begin{equation}
	n(\boldsymbol{\Omega})=\Psi^{\dagger}(\boldsymbol{\Omega})  \Psi(\boldsymbol{\Omega})=\frac{1}{N_s}\sum_{m_1, m_2} \bar{Y}_{s, m_1}^{\left(s\right)}(\boldsymbol{\Omega}) Y_{s, m_2}^{\left(s\right)}(\boldsymbol{\Omega}) \boldsymbol{c}_{m_1}^{\dagger} \boldsymbol{c}_{m_2},
\end{equation}
where we have used the monopole harmonics $Y_{s, m}^{\left(s\right)}(\boldsymbol{\Omega})$ derived above. The angular momentum decomposition of the total density operator is given by 
\begin{equation}
\begin{aligned}
	n_{l, m}& =(-1)^l \int d \boldsymbol{\Omega} Y_{l, m}(\boldsymbol{\Omega}) n(\boldsymbol{\Omega}) \\
	& = \sqrt{\frac{2 l+1}{4 \pi}} \sum_{m_1, m_2}(-1)^{3 s+m_1}\left(\begin{array}{ccc}
		s & l & s \\
		-s & 0 & s
	\end{array}\right)\left(\begin{array}{ccc}
		s & l & s \\
		-m_1 & m & m_2
	\end{array}\right) \boldsymbol{c}_{m_1}^{\dagger} \boldsymbol{c}_{m_2}.
\end{aligned}
\end{equation}
Here, $\boldsymbol{c}_{m}=(c_{m,1\uparrow}, c_{m,1\downarrow}, c_{m,2\uparrow}, c_{m,2\downarrow})^T$. Here $1(2)$ labels the layer index and $\uparrow (\downarrow)$ represents the spin index. $Y_{l,m}$ is the spherical harmonics. And the angular momentum decomposition of the local density operator of layer-$\tau$ is
\begin{equation}
\vec{n}_{l, m}^{\tau} = \sqrt{\frac{2 l+1}{4 \pi}} \sum_{m_1, m_2}(-1)^{3 s+m_1}\left(\begin{array}{ccc}
	s & l & s \\
	-s & 0 & s
\end{array}\right)\left(\begin{array}{ccc}
	s & l & s \\
	-m_1 & m & m_2
\end{array}\right) \boldsymbol{c}_{m_1,\tau}^{\dagger}\boldsymbol{\sigma} \boldsymbol{c}_{m_2,\tau}.
\end{equation}
Throughout this paper, the notation $\sigma^0$ and $\boldsymbol{\sigma}$ means $I$ and $(\sigma_x, \sigma_y, \sigma_z)$.

For any two-body interaction potential $U(\boldsymbol{\Omega}_{a, b})$ depending on $\boldsymbol{\Omega}_{a, b}$ on the spherical geometry, it can be expanded in Legendre polynomials
\begin{equation}
U(\boldsymbol{\Omega}_{a, b})=\sum_l U_l P_l\left(\boldsymbol{\Omega}_{a, b}\right)=\sum_{l, m} U_l \frac{4 \pi}{2 l+1} \bar{Y}_{l, m}\left(\boldsymbol{\Omega}_a\right) Y_{l, m}\left(\boldsymbol{\Omega}_b\right).
\end{equation}
Taking the first term in the $H_{\mathrm{int}}$ as an example, we can obtain its second-quantization form is 
\begin{equation}
\begin{aligned}
	H_{\mathrm{int}}^{0}&=\int d \boldsymbol{\Omega}_{a, b}U_0 n\left(\boldsymbol{\Omega}_a\right) n\left(\boldsymbol{\Omega}_b\right)
	&=\int d \boldsymbol{\Omega}_{a, b} \sum_{l, m} U^0_l \frac{4 \pi}{2 l+1} \bar{Y}_{l, m}\left(\boldsymbol{\Omega}_a\right)n\left(\boldsymbol{\Omega}_a\right) Y_{l, m}\left(\boldsymbol{\Omega}_b\right)n\left(\boldsymbol{\Omega}_b\right)
	&=\sum_{l, m} U^0_l \frac{4 \pi}{2 l+1}n^\dagger_{l, m}n_{l, m}.
\end{aligned}
\end{equation}
Similar form can be found for the second and third term. Finally, our second quantization form of the real-space Hamiltonian $H_{\mathrm{int}}$ is given by 
\begin{equation}
	\begin{aligned}
		H_{\mathrm{int}} &=\sum_l \frac{4 \pi}{2 l+1} U^0_l \sum_m n^\dagger_{l, m}n_{l, m}-\sum_l \frac{4 \pi}{2 l+1} U_l^1 \sum_m\left(\vec{n}_{l, m}^{1\dagger} \vec{n}_{l, m}^1+\vec{n}_{l, m}^{2\dagger} \vec{n}_{l, m}^2\right) \\
		& +\sum_l \frac{4 \pi}{2 l+1} U_l^2 \sum_m\vec{n}_{l, m}^{1\dagger} \vec{n}_{l, m}^2-h \sum_m \boldsymbol{c}_m^{\dagger} \tau^x \sigma^0 \boldsymbol{c}_m.
	\end{aligned}
\end{equation}

\subsection{A. 3. Relationship with the Haldane pseudopotential}
The real-space interactions we used in the main text are connected to the Haldane pseudopotential $V_l$. For the cases of $\delta$ potential and its derivative, the only non-zero component of the Haldane pseudopotential is $l=0$ and $l=1$, respectively. Similarly, we can use the Haldane pseudopotential to reformulate the Hamiltonian 
\begin{equation}
H_{\mathrm{int}}^{0}=\sum_{m_1 m_2 m_3 m_4}V_{m_1 m_2 m_3 m_4}\left(\hat{\mathbf{c}}_{m_1}^{\dagger} \hat{\mathbf{c}}_{m_4}\right)\left(\hat{\mathbf{c}}_{m_2}^{\dagger} \hat{\mathbf{c}}_{m_3}\right),
\end{equation}
and the parameter $V_{m_1 m_2 m_3 m_4}$
is connected to the Haldane pseudopotential $V_l$ by
\begin{equation}
V_{m_1 m_2 m_3 m_4}=\sum_l V_l(4 s-2 l+1)\left(\begin{array}{ccc}
s & s & 2 s-l \\
m_1 & m_2 & -m_1-m_2
\end{array}\right)\left(\begin{array}{ccc}
s & s & 2 s-l \\
m_4 & m_3 & -m_4-m_3
\end{array}\right).
\end{equation}
By comparing the coefficients, we can get the relations between $V_l$ and $U_l$,
\begin{equation}
V_{2s-l}=(-1)^{2s+l} \sum_{k} U_{k}\left(\begin{array}{ccc}
s & k & s \\
-s & 0 & s
\end{array}\right)^2\left\{\begin{array}{ccc}
s & s & l \\
s & s & k
\end{array}\right\}.
\end{equation}
Here, $\left\{\begin{array}{ccc}
s & s & l \\
s & s & k
\end{array}\right\}$ is Wigner $6j$ coefficient.

\subsection{A.4 Connecting with the $\mathrm O(N)$ models}
In the main text, we have shown a microscopic realization of Heisenberg-like model exhibiting Wilson-Fisher $\mathrm O(3)$ transition. Here we discuss that  
the proposed fuzzy sphere bilayer model has far-reaching extension to realize general universality class of Wilson-Fisher  $\mathrm O(N)$ transition with $N=1$ (Ising), $N=2$ (XY) and $N=3$ (Heisenberg).

To clarify the construction, we rewrite the density-density interaction in each layer as (The interlayer interaction doesnot specify here)	
\begin{align}
H_{\textrm{spin},\tau}&= - \int\!d\vec{\Omega}_{a,b} [U_1 (n^x_\tau(\vec{\Omega}_a) \cdot n^x_\tau(\vec{\Omega}_b) +
 n^y_\tau(\vec{\Omega}_a) \cdot n^y_\tau(\vec{\Omega}_b) +
 \Delta^z n^z_\tau(\vec{\Omega}_a) \cdot n^z_\tau(\vec{\Omega}_b) )]
\end{align}
The definition of local density operators are in line with the main text:
the local density operator of layer-$\tau$ is $\vec{n}_\tau(\vec \Omega) = (n_\tau^x,n_\tau^{y},n_\tau^z) =\psi^\dagger_\tau(\vec \Omega) \vec{\sigma} \psi_\tau(\vec \Omega)$. 
Here setting $\Delta^z=1$ favors the Heisenberg $\mathrm O(3)$ symmetry as discussed in the main text, which preserves the spin rotation symmetry. We refer $\Delta^z=1$ as Heisenberg point in the following discussion. 
Further tuning the parameter $\Delta^z$ leads to different global symmetry of effective spin degrees of freedom: 
1) $\Delta^z>1$ (``easy-axis'' condition) leads to a Ising $Z_2$ symmetry;
2) $\Delta^z<1$ (``easy-plane'' condition) leads to a XY $\mathrm O(2)$ symmetry. 

Moreover, we should be careful on the form of interactions.
For the Heisenberg model in the main text, we choose short-ranged interactions to be 
$U_1=u_1\delta(\vec{\Omega}_1-\vec{\Omega}_2)$. This form of interaction is special to achieve $\mathrm O(3)$ spin, because $\delta(\vec{\Omega}_1-\vec{\Omega}_2)$ potential is always SU(2) symmetric. To realize the XY transition and Ising transition, one also needs to modify the form of density-density interaction. In specific, we use the following form:
\begin{align}
&U_1 = u_1^0 \delta(\vec{\Omega}_1-\vec{\Omega}_2) + u^1_1 \nabla^2 \delta(\vec{\Omega}_1-\vec{\Omega}_2) 
\end{align}
Tuning the parameters $\Delta^z$ and $u^0_1, u^1_1$ breaks the spin $\mathrm O(3)$ rotation symmetry and
produce XY spin or $Z_2$ spin.
To demonstrate it, we compute the energy spectrum of a single layer model as a function of $\Delta^z$, in Fig. \ref{sfig:delta}. At the Heisenberg point, the ground state is $2S+1$-fold degenrated guaranteed by the spin rotational symmetry for quantum spins. For $\Delta^z>1$, the ground state is doublet degenerated, related to the $Z_2$ symmetric breaking Ising spin. For $\Delta^z<1$, the ground state belongs to the XY spin order.

\begin{figure}[!htb]
	\includegraphics[width=0.65\textwidth]{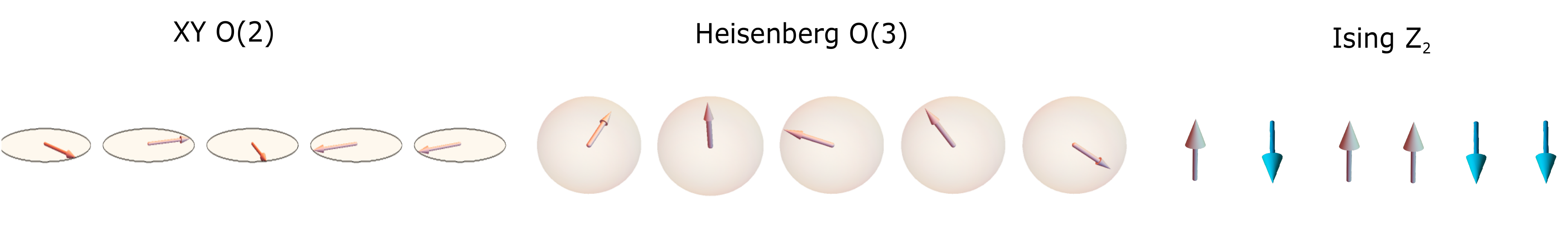}
 \includegraphics[width=0.35\textwidth]{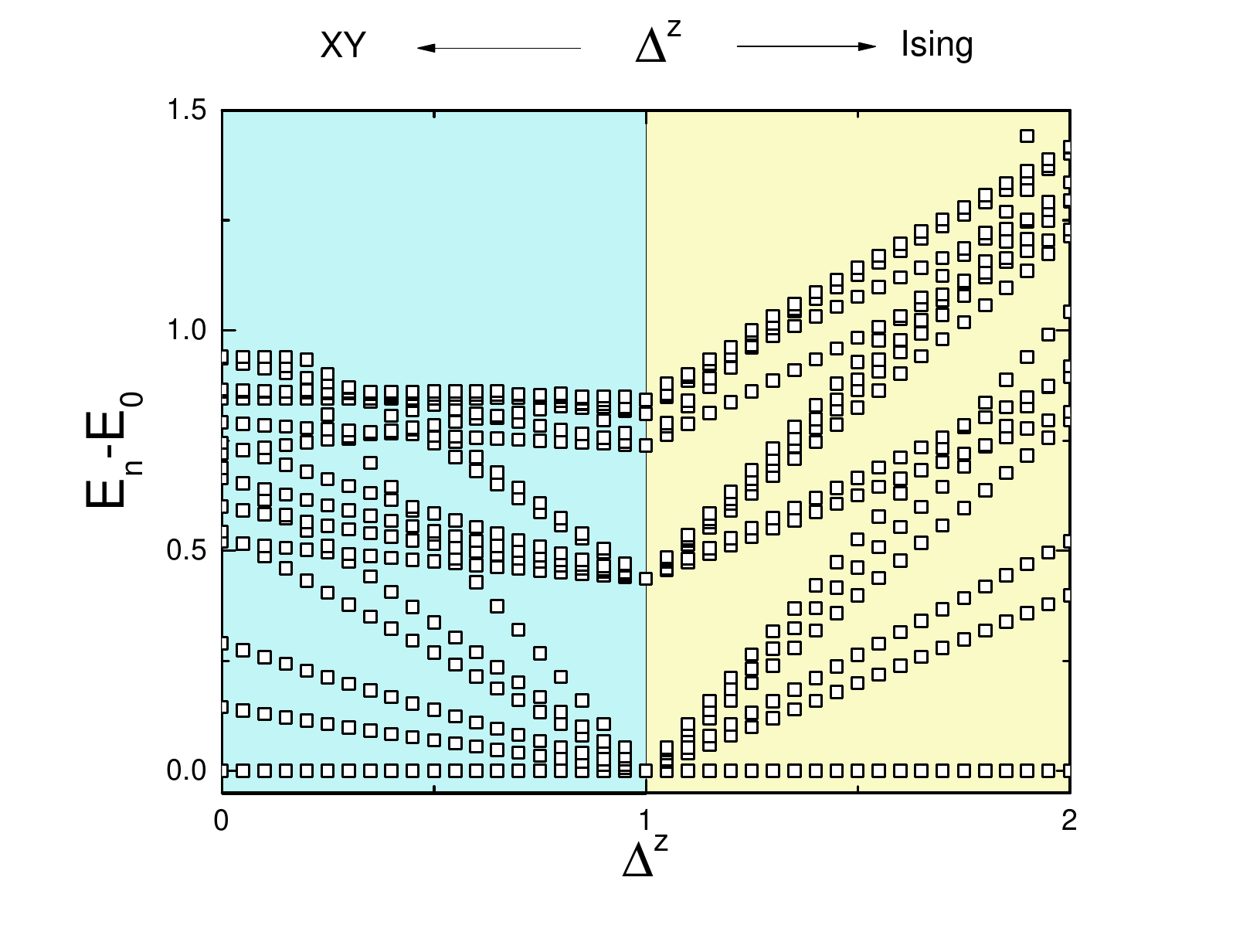}
	\caption{(Top panel) Fuzzy sphere model can simulate magnetic order with $\mathrm O(N)$ symmetry: $N=1$ (Ising), $N=2$ (XY) and $N=3$ (Heisenberg). 
	(Bottom panel) Energy spectra of $H_{\mathrm{spin},\tau}$ as a function of $\Delta^z$. At the Heisenberg point $\Delta^z=1$, the ground state is $2S+1$-fold degenerated. By tuning $\Delta^z$ away from the Heisenberg point leads to quantum XY spins or quantum Ising spins. }
	\label{sfig:delta}
\end{figure}

Based on the magnetic orders with different symmetries established on the fuzzy sphere bilayer model, one can further construct phase transitions belong to Wilson-Fisher $\mathrm O$(N) universality class: $N=1$ (Ising), $N=2$ (XY) and $N=3$ (Heisenberg). The Ising transition should be in line with  the existing discussion \cite{ZHHHH2022}, and the Heisenberg transition has been studied in the main text. We believe that the XY transition can be also realized in the fuzzy sphere bilayer model in a similar way.

Finally, let us make some remarks here. Previously we have realized a 3D Ising transition on the fuzzy sphere model with one-layer \cite{ZHHHH2022} (We dubbed it as single-layer fuzzy sphere model). 
The idea is to use a spin-flip transverse field term (i.e. involving $n^x(\Omega)$ operator) to induce a paramagnetic phase. Unfortunately, this single-layer fuzzy sphere model cannot realize the XY or Heisenberg transition.  The reason is, the spin-flip transverse field term involving operator explicitly breaks the spin rotation symmetry, so that a $O(2)$ or $O(3)$ symmetric paramagnet cannot be created on the single-layer fuzzy sphere model.
Here to overcome this obstacle, we turn to construct a fuzzy sphere bilayer model, which is able to realize the $\mathrm O$(N) transition with global continuous symmetry.  In this context, the fuzzy sphere bilayer model is meaningful, which goes beyond the limitation of single-layer fuzzy sphere model.

\section{B. Physical observables across the phase transition}

In the main text, the critical point of the phase transition is determined by the scaling of local order parameter.
In this section, we consider two more physical observables, the binder cumulant $U_4$ and the lowest energy gap, and study their finite-size scaling around the phase transition to confirm the estimation of $h_c$. In the end, we calculate the charge gap to show that spin is the degree of freedom in the low-energy region.

The binder cumulant $U_4$ \cite{Hasenbusch2023} is defined as
\begin{equation}
U_4=\langle M_\tau^4\rangle/\langle M_\tau^2\rangle^2.
\end{equation}
It is a universal quantity related to the four-point field $\phi$ at the phase transition. Fig.~\ref{fig:U4} (a) shows $U_4$ with respect to the transverse field strength $h$ for different $N_s$. Clearly, at small $h$ the model is the ordered Heisenberg magnet, while at large $h$ the model is the disordered paramagnet. There is a crossing region $h\approx h_c$, where different system sizes cross with each other. The precise value of critical point $h_c \approx 0.2167 \pm 0.0002$ can be determine by the crossing-point analysis according to the scaling form $h_c(N_s)=aN_s^{-(1/\nu+\omega)/2}+b$, shown in Fig.~\ref{fig:U4} (a) (inset). 
This is the same analysis as how we used the order parameter in the main text, while the critical point is a little smaller. 

Secondly, we compute the energy gap $\Delta$ between the ground state and the first excited state for different system sizes $N_s=4-8$, and do finite-size scaling. As one can find in Fig.~\ref{fig:U4} (b), below $h=0.232$, the energy gap $\Delta$ is smaller than $0$, while it becomes gapped when $h$ exceeds $0.232$, which gives rise to $h_c \approx 0.232$. Three different physical observables, including the order parameter in the main text, are largely consistent with each other.

\begin{figure}[!b]
	\includegraphics[width=0.45\textwidth]{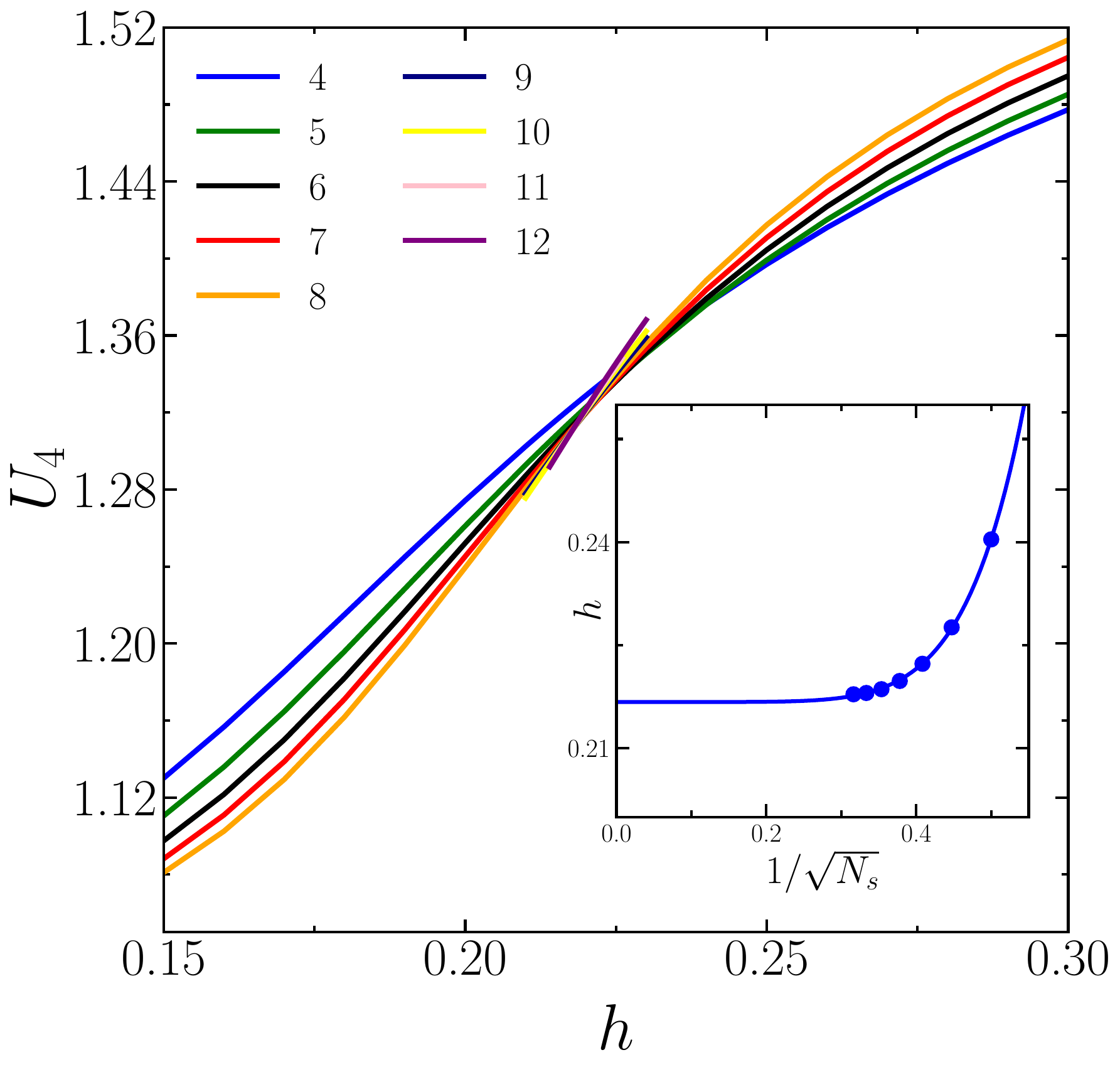}
	\caption{The binder cumulant $U_4$ with different system sizes $N_s=4-12$. (Inset) Finite-size scaling of crossing points by a finite-size pair $(N_s,N_s+1)$, which gives rise to $h_c \approx 0.2167 \pm 0.0002$. The scaling form is $h_c(N_s)=aN_s^{-(1/\nu+\omega)/2}+b$.
 }
	\label{fig:U4}
\end{figure}

\begin{figure}[!b]
    \includegraphics[width=0.45\textwidth]{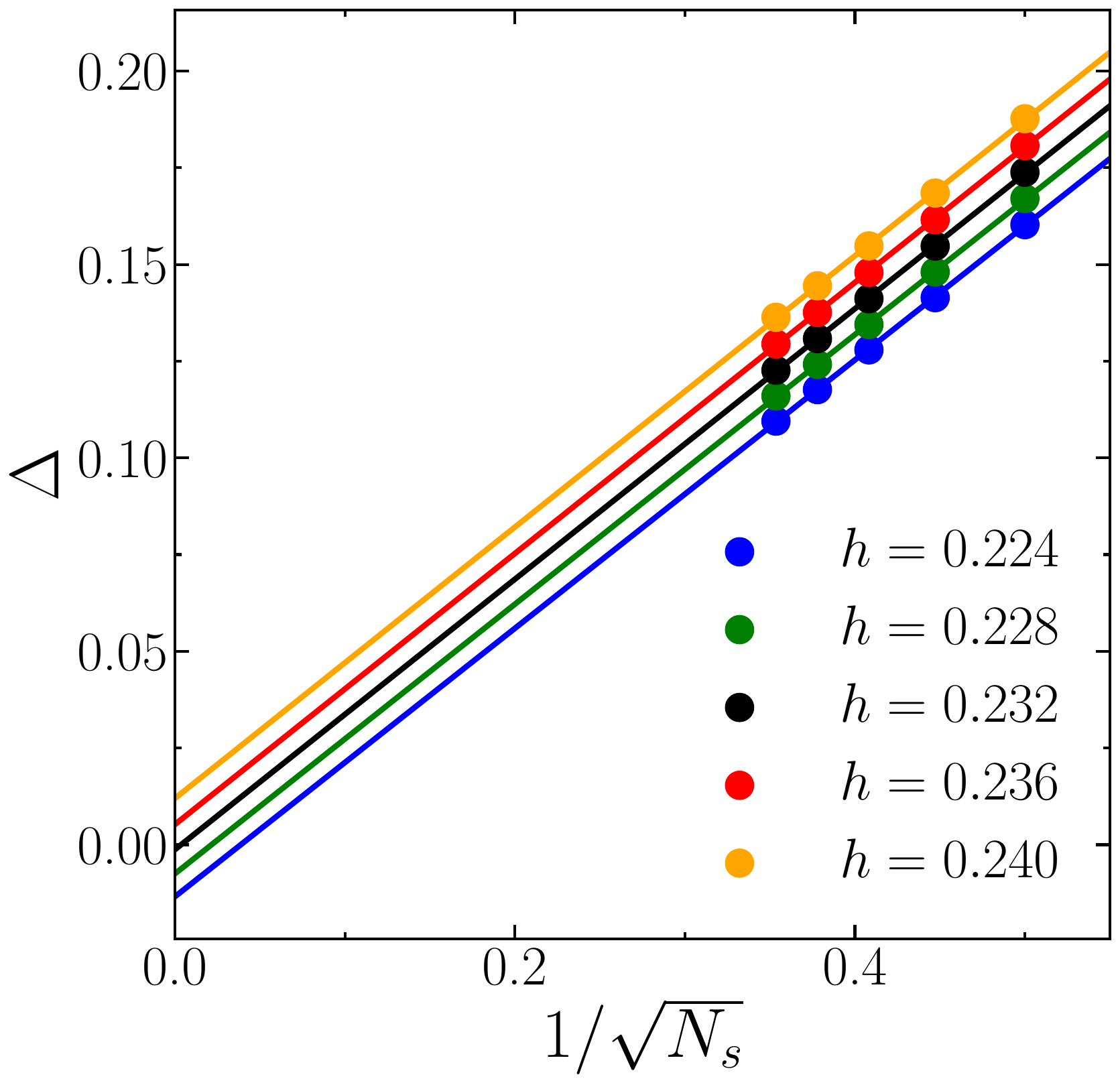}
	\caption{
 (b) Finite-size scaling of the lowest energy gap with different $h$ for various $N_s=4-8$, which gives us the critical point $h_c \approx 0.232$.
 }
	\label{fig:gap_scaling}
\end{figure}

Thirdly, the original fuzzy sphere model consists of fermionic degrees of freedom. So one preliminary question to be addressed in the model design is whether or not the charge excitation gap is relevant to the discussion of magnetic phase transition. Here we define the charge gap as $\Delta_c=E_0(N+1, N_s)+E_0(N-1, N_s)-2E_0(N, N_s)$, where $E_0(N, N_s)$ is the ground state energy on $N_s$ lowest Landau level orbitals filled by $N$ electrons. After obtaining the charge gap on each system size, we perform a finite-size scaling to estimate the charge gap in the thermodynamic limit. As shown in Fig.~\ref{fig:ChargeGap}, the charge gap at the critical point $h=h_c$ is nonzero on all system sizes, and the value in the thermodynamic limit is also finite. Thus, we conclude that the spin degrees of freedom undergoes a phase transition while the charge degrees of freedom are always gapped. And close to the phase transition point the spin excitation, rather than the charge excitation, dominates the low-energy excitation. 

\begin{figure}
	\includegraphics[width=0.35\textwidth]{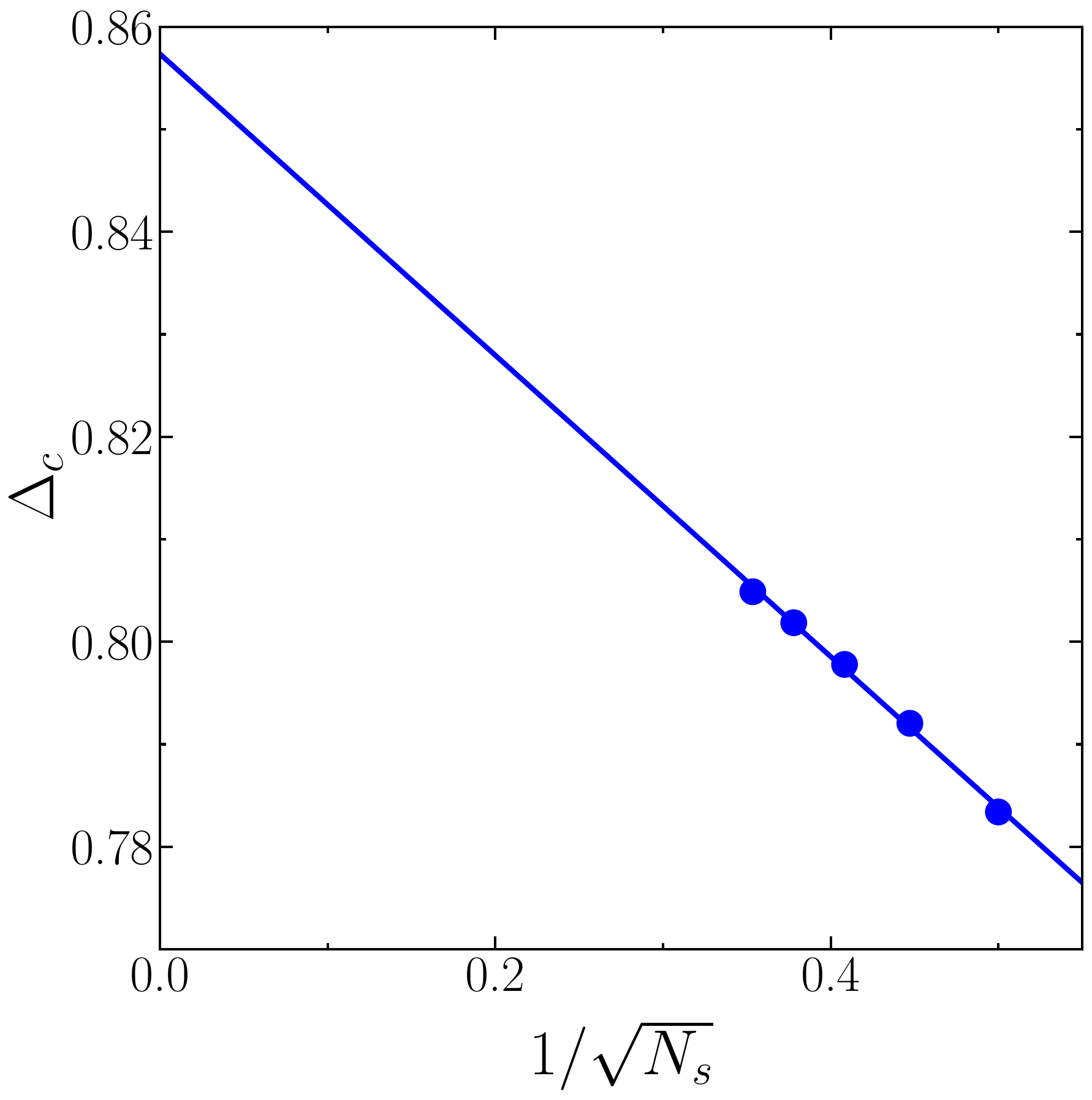}
	\caption{Finite-size scaling of charge excitation gap with system sizes $N_s=4-8$ at the phase transition point $h_c$.}
	\label{fig:ChargeGap}
\end{figure}

\section{C. OPE coefficients}
This section delves into the nuances of calculating OPE coefficients, emphasizing the tensor structures and accounting for finite-size corrections.
\subsection{C.1. Tensor structures}
In our main text, we have analyzed five OPE coefficients, namely $f_{\phi\phi s},f_{sss},f_{\phi\phi T},f_{ss T}$ and $f_{\phi\phi J_{\mu}}$. These coefficients correspond to four primary fields: \(\phi\), \(s\), \(T_{\mu\nu}\), and \(J_\mu\). Here, we outline their tensor structures:
\begin{align}\nonumber
    \langle \phi_i(x_1) \phi_j(x_2) s(x_3) \rangle &= \delta_{ij}\frac{f_{\phi\phi s}}{x_{12}^{2\Delta_\phi-\Delta_s} x_{23}^{\Delta_s} x_{13}^{\Delta_s}} \nonumber \\
    \langle \phi_i(x_1) \phi_j(x_2) T(x_3,z) \rangle & = \delta_{ij}
    \frac{f_{\phi\phi T}\left(\frac{\bm x_{13}\cdot \bm z}{x_{13}^2}-\frac{\bm x_{23}\cdot \bm z}{x_{23}^2}\right)^2}
    {x_{12}^{2\Delta_\phi-\Delta_T+2} x_{23}^{\Delta_T-2} x_{13}^{\Delta_T-2}}
    \nonumber \\
    \langle s(x_1) s(x_2) s(x_3) \rangle &=\frac{ f_{sss}}{x_{12}^{\Delta_s} x_{23}^{\Delta_s} x_{13}^{\Delta_s}} \nonumber \\
    \langle s(x_1) s(x_2) T(x_3,z) \rangle & = 
    \frac{f_{ss T}\left(\frac{\bm x_{13}\cdot \bm z}{x_{13}^2}-\frac{\bm x_{23}\cdot \bm z}{x_{23}^2}\right)^2}
    {x_{12}^{2\Delta_s-\Delta_T+2} x_{23}^{\Delta_T-2} x_{13}^{\Delta_T-2}}
    \nonumber \\
    \langle \phi_i(x_1) \phi^{\dagger j}(x_2) J^A(x_3,z) \rangle & = -T^A_{~ij}
    \frac{f_{\phi\phi J}\left(\frac{\bm x_{13}\cdot \bm z}{x_{13}^2}-\frac{\bm x_{23}\cdot \bm z}{x_{23}^2}\right)}
    {x_{12}^{2\Delta_\phi-\Delta_J+1} x_{23}^{\Delta_J-1} x_{13}^{\Delta_J-1}} 
\end{align}
Where $z$ serves as the auxiliary coordinate, devoid of indices, pertinent to the Lorentz tensor. The indices $i$, $j$, and $A$ relate to the global $O(3)$ structure. Within the OPE linked to the current operator, the tensor $T^A_{~ij}$ functions as the $O(3)$ generators. It's crucial to recognize that the tensor structures of these OPEs use the two-point correlators for normalization.
\begin{align}\nonumber
    \langle \phi_i(x)\phi_i(0) \rangle &= \frac{\delta_{ij}}{x^{2\Delta_\phi}}
    \nonumber \\
    \langle T(x,z_1)T(0,z_2)\rangle &= \frac{\left(\frac12
    \bm z_1\cdot \bm z_2-(\bm n\cdot \bm z_1)(\bm n\cdot \bm z_2)
    \right)^2}{x^{2\Delta_T}} \nonumber \\
    \langle s(x)s(0) \rangle &= \frac{1}{x^{2\Delta_s}}
    \nonumber \\
    \langle J^A(x,z_1)J^B(0,z_2)\rangle &= \tau^{AB}\frac{\left(\frac12
    \bm z_1\cdot \bm z_2-(\bm n\cdot \bm z_1)(\bm n\cdot \bm z_2)
    \right)}{x^{2\Delta_J}}
\end{align}
Here, $\bm n = \frac{\bm x}{x}$ denotes the unit vector and $\tau^{AB}=\mathrm{Tr}[T^A T^b]=2\delta^{AB}$. To recover the explicit indices, one can apply the differential operator relative to the auxiliary coordinate $z$:
\begin{equation}
    D_\mu = \frac{D-2}{2}\frac{\partial}{\partial z^\mu}
    - z_\nu\frac{\partial^2}{\partial z_\mu\partial z^\mu}
    -\frac12 z_\mu\frac{\partial^2}{\partial z_\nu\partial z^\nu}.
\end{equation}
The operator, exhibiting explicit Lorentz quantum numbers $\ell$, $m$ alongside $O(3)$ quantum numbers $s$, $s_z$, can be formulated as:
\begin{equation}
    \mathcal{O}_{\ell,m}^{s,s_z}(x) = n_{(\ell,m)}^{\mu_1\cdots\mu_\ell}e^{(s,s_z)}_{ij\cdots} \mathcal{O}^{ij\cdots}_{\mu_1\cdots\mu_\ell}(x)
\end{equation}
where $ n_{(\ell,m)}^{\mu_1\cdots\mu_\ell}$ and $e^{(s,s_z)}_{ij\cdots} $ are Lorentz and $O(3)$ polarization respectively. By leveraging the state-operator correspondence, this operator transforms into the state within the numerical states, subject to normalization
\begin{equation}
    |\mathcal O^{s,s_z}_{\ell,m}\rangle = N(\mathcal O^{s,s_z}_{\ell,m})\lim_{x\rightarrow 0} \mathcal O^{s,s_z}_{\ell,m}(x)|0\rangle.
\end{equation}
Thus
\begin{equation}
    N(\mathcal O^{s,s_z}_{\ell,m}) = \left[
    \lim_{x\rightarrow\infty}x^{2\Delta_\mathcal{O}} 
    \left( n^* \right)^{\mu_1^\prime\cdots\mu_\ell^\prime}_{\ell,m}
     n^{\mu_1\cdots\mu_\ell}_{\ell,m}
    \left( e^* \right)_{i^\prime j^\prime\cdots}^{s,s_z}
    e_{i j\cdots}^{s,s_z}
    I_{\mu_1^\prime}^{~~\nu_1}(x)\cdots I_{\mu_\ell^\prime}^{~~\nu_\ell}(x)
    \langle \mathcal{O}^{i^\prime j^\prime\cdots}_{\nu_1\cdots \nu_\ell}(x)\mathcal{O}^{i j\cdots}_{\mu_1\cdots \mu_\ell}(0) \rangle
    \right]^{-1/2}
\end{equation}
where $I_{\mu_1}^{~~\nu_1}(x)=\delta_\mu^{~~\nu}-2n_\mu n_\nu$ comes from the conjugation.
For a deeper insight, let's examine the tensor structure of the OPE $\phi\phi J$ as a representative example. For comprehensive tensor structures associated with other OPEs, readers are directed to earlier works\cite{Hu2023, zhou2023so5}. 
The correlators related to $\phi$ and $J$ are
\begin{equation}
\begin{split}
    \langle \phi_i(x)\phi_j(0) \rangle &= \delta_{ij} x^{-2\Delta_\phi}\\
    \langle J^A_\mu(x)J^B_\nu(0) \rangle &= \frac14 \delta^{AB} \left(\eta_{\mu\nu}-2n_\mu n_\nu\right)x^{-2\Delta_J}
\end{split}
\end{equation}
Both the Lorentz and $O(3)$ polarizations serve as spherical tensors for the $SO(3)$ rotation group. Therefore:
\begin{align}
    e^{(0,0)} = n_{(0,0)}&=1&
    e^{(1,0)}_z = n_{(1,0)}^z&=1\nonumber\\
    e^{(1,\pm 1)}_x = n_{(1,\pm 1)}^x&=\mp 1/\sqrt{2} &e^{(1,\pm 1)}_y =n_{(1,\pm 1)}^y&=- i/\sqrt{2}
\end{align}
Consequently, the normalization factors for $\phi$ and $J$ are:
\begin{align}
    &N\left(\phi^{(1,\pm 1)}_{(0,0)}\right) =N\left(\phi^{(1,0)}_{(1,0)}\right)= 1 &N\left(J^{(1,1)}_{1,0}\right) = 2.
\end{align}

Following the same procedure, the OPE of $\phi\phi J$ in $\mathbb S^2\times \mathbb R$ is:
\begin{equation}
\begin{split}
    \langle\phi^{(1,1)}_{(0,0)}|\phi^{(1,0)}(\vec{\Omega})|J^{(1,1)}_{(1,0)}\rangle&=
    N^*\left(\phi^{(1,\pm 1)}_{(0,0)}\right)
    N\left(J^{(1,1)}_{1,0}\right)
    \lim_{x\to\infty}x^{2\Delta_\phi}
    (e^*)^{(1,1)}_{i}e^{(1,0)}_{j}e^{(1,1)}_{A}n^\mu_{(1,0)}
    \left\langle\phi^i(x)\phi^j(\Omega)J^A_\mu(0)\right\rangle\\
    & = R^{-\Delta_\phi}f_{\phi\phi J}\cos\theta
\end{split}
\end{equation}
By integrating the angle dependence and taking into account the angular momentum component
\begin{equation}
    \mathcal O_{l,m} = \int \mathrm d \vec{\Omega} \bar Y_{l,m}(\vec{\Omega})\mathcal{O}(\vec{\Omega})
\end{equation}
and the factor $R^{-\Delta_\phi}$ can be offset by its two-point correlator
\begin{equation}
    \langle \phi^{(1,0)}_{(0,0)}|\phi^{(1,0)}(\vec{\Omega})|0\rangle = R^{-\Delta_\phi}.
\end{equation}
In conclusion, the value of $f_{\phi\phi J}$ is:
\begin{equation}
    f_{\phi\phi J} = \sqrt 4 \frac{\int\mathrm d \vec{\Omega} \bar Y_{1,0}(\vec{\Omega})\langle\phi^{(1,1)}_{(0,0)}|\phi^{(1,0)}(\vec{\Omega})|J^{(1,1)}_{(1,0)}\rangle}{\int\mathrm d \Omega \bar Y_{0,0}(\vec{\Omega})\langle \phi^{(1,0)}_{(0,0)}|\phi^{(1,0)}(\vec{\Omega})|0\rangle}
     = \sqrt 3 \frac{\langle\phi^{(1,1)}_{(0,0)}|\phi^{(1,0)}_{(1,0)}|J^{(1,1)}_{(1,0)}\rangle}{\langle \phi^{(1,0)}_{(0,0)}|\phi^{(1,0)}_{(0,0)}|0\rangle}.
\end{equation}
Now we discuss about other OPEs. Since they are both scalar-scalar-scalar(spin-$\ell=2$ tensor) type, we can directly use the results in Ref.\cite{Hu2023,zhou2023so5} after treating the scalar $\phi_i$ in the vector representation of $O(3)$. If we choose the $s=1,s_z=0$ component of $\phi_i$, the polarization is just $e^{(1,0)}_z=1$ and will not produce any modification. Finally, we have
\begin{equation}
\begin{split}
    f_{\phi\phi s} &= 
     \frac{\langle\phi^{(1,0)}_{(0,0)}|\phi^{(1,0)}_{(0,0)}|s^{(0,0)}_{(0,0)}\rangle}{\langle \phi^{(1,0)}_{(0,0)}|\phi^{(1,0)}_{(0,0)}|0\rangle} \\
    f_{sss} &= 
     \frac{\langle s^{(0,0)}_{(0,0)}|s^{(0,0)}_{(0,0)}|s^{(0,0)}_{(0,0)}\rangle}{\langle s^{(0,0)}_{(0,0)}|s^{(0,0)}_{(0,0)}|0\rangle} \\
    f_{\phi\phi T} &= \sqrt\frac{15}{8}
    \frac{\langle\phi^{(1,0)}_{(0,0)}|\phi^{(1,0)}_{(2,0)}|T^{(0,0)}_{(2,0)}\rangle}{\langle \phi^{(1,0)}_{(0,0)}|\phi^{(1,0)}_{(0,0)}|0\rangle}\\
    f_{ssT} &= \sqrt\frac{15}{8}
     \frac{\langle s^{(0,0)}_{(0,0)}|s^{(0,0)}_{(2,0)}|T^{(0,0)}_{(2,0)}\rangle}{\langle s^{(0,0)}_{(0,0)}|s^{(0,0)}_{(0,0)}|0\rangle}
\end{split}
\end{equation}
\subsection{C.2. Relation to central charge}
The relation between OPE coefficients and the central charge is evident from the study of their correlators. Specifically, the OPE coefficients, denoted by \(\mathcal O\mathcal O T\) and \(\mathcal O\mathcal O J\), can be associated with the central charge \(C_T\) and the current central charge \(C_J\). These relationships are determined by the correlators:
\begin{equation}
\begin{split}
    \langle T(x,z_1)T(0,z_2)\rangle &= C_T\frac{\left( \frac 12
    \bm  z_1\cdot \bm z_2-(\bm n\cdot \bm z_1)(\bm n\cdot \bm z_2)
    \right)^2}{x^{2\Delta_T}} \\
    \langle J^A(x,z_1)J^B(0,z_2)\rangle &= C_J\tau^{AB}\frac{\left(
    \frac12 \bm z_1\cdot \bm z_2-(\bm n\cdot \bm z_1)(\bm n\cdot \bm z_2)
    \right)}{x^{2\Delta_J}}
\end{split}
\end{equation}
By referencing Ref.\cite{RevModPhys.91.015002}, we find the CB prediction values of the central charge:
\begin{equation}
\begin{split}
    f_{\phi\phi T} &= \frac{3\Delta_\phi}{4\sqrt{C_T}}\approx 0.1889  \\
    f_{ss T} &= \frac{3\Delta_s}{4\sqrt{C_T}}\approx 0.5805 \\
    f_{\phi\phi J} &= \frac{1}{\sqrt{C_J}}\approx 0.7428.
\end{split}
\end{equation}
These approximations utilize bootstrap values from Ref.\cite{Su2021, RevModPhys.91.015002}, where \(C_T/C_T^{\text{free}}=0.9445\) and \(C_J/C_J^{\text{free}}=0.9063\). Additionally, theoretical values cited from Ref.\cite{PETKOU1995101} provide \(C_T^{\text{free}} = N\frac{d}{d-1}=4.5\) and \(C_J^{\text{free}}=\frac{2}{d-2}=2\).

\subsection{C.3. Finite size corrections}
In this section, we will provide a thorough finite-size scaling of OPE coefficients \cite{Hu2023} from the microscopic spin operators.

Since the lowest $O(3)$ vector $\phi$ corresponds to the local order parameter, and the lowest scalar $s=\phi^2$ in the Wilson-Fisher description using quantum field theory. We will choose local operator $\vec{n}_\tau(\vec{\Omega})$ to approach the CFT operator $\phi$, and $\vec{n}_\tau^2(\vec{\Omega})$ to approach $s$. The operator decomposition $\vec{n}_\tau(\vec{\Omega})$ generically is,
\begin{equation}\label{eq:phi_decomposition}
\vec{n}_\tau(\boldsymbol{\Omega})=c_\phi \phi(\boldsymbol{\Omega})+c_{\partial_\mu \phi} \partial_\mu \phi(\boldsymbol{\Omega})+c_{\square \phi} \square \phi(\boldsymbol{\Omega})+c_{\partial_\mu \partial_\nu \phi} \partial_\mu \partial_\nu \phi(\boldsymbol{\Omega})+\cdots
\end{equation}
where the first four terms represents the primary $\phi$ and components of its descendants. More other descendants of $\phi$, and other primaries and corresponding descendants, included in $\cdots$. Similarly, the operator $\vec{n}^2_\tau(\vec{\Omega})$ can be disassembled by
\begin{equation}
\vec{n}^2_\tau(\boldsymbol{\Omega})=c_I I+\left[c_s s(\boldsymbol{\Omega})+c_{\partial_\mu s} \partial_\mu s(\boldsymbol{\Omega})+\cdots\right]+\cdots.
\end{equation}
Two scaling dimensions are $\Delta_\phi\approx 0.519$ and $\Delta_s\approx 1.595$, respectively. Now, we can extract the OPE coefficient $f_{\phi\phi s}$ by $\frac{\left\langle\phi^{(1,0)}_{(0,0)}\left|[\boldsymbol{n}_\tau]^{(1,0)}_{(0,0)}\right| s^{(0,0)}_{(0,0)}\right\rangle}{\left\langle\phi^{(1,0)}_{(0,0)}\left|[\boldsymbol{n}_\tau]^{(1,0)}_{(0,0)}\right| 0\right\rangle}$,
for which only operators with same quantum number will contribute,
\begin{equation}\label{eq:phiphis}
\begin{aligned}
\frac{\left\langle\phi^{(1,0)}_{(0,0)}\left|[\boldsymbol{n}_\tau]^{(1,0)}_{(0,0)}\right|s^{(0,0)}_{(0,0)}\right\rangle}{\left\langle\phi^{(1,0)}_{(0,0)}\left|[\boldsymbol{n}_\tau]^{(1,0)}_{(0,0)}\right| 0\right\rangle}\approx &
\frac{ c_\phi f_{\phi \phi s} R^{-\Delta_\phi}+c_{\square \phi} f_{\phi, \square \phi, s} R^{-\left(\Delta_\phi+2\right)}+c_{\square^2 \phi} f_{\phi, \square^2 \phi, s} R^{-\left(\Delta_\phi+4\right)}+\cdots}
 {c_\phi R^{-\Delta_\phi}+c_{\square \phi} R^{-\left(\Delta_\phi+2\right)}+c_{\square^2 \phi} R^{-\left(\Delta_\phi+4\right)}+\cdots}\\
 \approx & f_{\phi \phi s}+\frac{c_1}{R^2}+\frac{c_2}{R^4}+O(R^{-6})\approx  f_{\phi \phi s}+\frac{c_1}{N_s}+\frac{c_2}{N_s^2}+O(N_s^{-3}).
\end{aligned}
\end{equation}
Similarly, $f_{\phi\phi s}$ can also be computed by 
\begin{equation}\label{eq:phisphi}
\begin{aligned}
\frac{\left\langle\phi^{(1,0)}_{(0,0)}\left|[\boldsymbol{n}^2_\tau]^{(0,0)}_{(0,0)}\right|\phi^{(1,0)}_{(0,0)}\right\rangle-\left\langle 0\left|[\boldsymbol{n}^2_\tau]^{(0,0)}_{(0,0)}\right| 0\right\rangle}{\left\langle s^{(0,0)}_{(0,0)}\left|[\boldsymbol{n}^2_\tau]^{(0,0)}_{(0,0)}\right| 0\right\rangle}\approx &
\frac{ c_s f_{\phi s\phi} R^{-\Delta_s}+c_{\square s} f_{\phi, \square s, \phi} R^{-\left(\Delta_s+2\right)}+c_{\square^2 s} f_{\phi, \square^2 s, \phi} R^{-\left(\Delta_s+4\right)}+\cdots}
 {c_s R^{-\Delta_s}+c_{\square s} R^{-\left(\Delta_s+2\right)}+c_{\square^2 s} R^{-\left(\Delta_s+4\right)}+\cdots}\\
 \approx & f_{\phi \phi s}+\frac{c_1}{R^2}+\frac{c_2}{R^4}+O(R^{-6}) \approx  f_{\phi \phi s}+\frac{c^\prime_1}{N_s}+\frac{c^\prime_2}{N_s^2}+O(N_s^{-3}),
\end{aligned}
\end{equation}
and the finite-size scaling of the OPE coefficient $f_{sss}$ reads
\begin{equation}
\begin{aligned}
\frac{\left\langle s^{(0,0)}_{(0,0)}\left|[\boldsymbol{n}^2_\tau]^{(0,0)}_{(0,0)}\right|s^{(0,0)}_{(0,0)}\right\rangle-\left\langle 0\left|[\boldsymbol{n}^2_\tau]^{(0,0)}_{(0,0)}\right| 0\right\rangle}{\left\langle s^{(0,0)}_{(0,0)}\left|[\boldsymbol{n}^2_\tau]^{(0,0)}_{(0,0)}\right| 0\right\rangle}\approx &
\frac{ c_s f_{sss} R^{-\Delta_s}+c_{\square s} f_{s, \square s, s} R^{-\left(\Delta_s+2\right)}+c_{\square^2 s} f_{s, \square^2 s, s} R^{-\left(\Delta_s+4\right)}+\cdots}
 {c_s R^{-\Delta_s}+c_{\square s} R^{-\left(\Delta_s+2\right)}+c_{\square^2 s} R^{-\left(\Delta_s+4\right)}+\cdots}\\
 \approx & f_{sss}+\frac{c_1}{R^2}+\frac{c_2}{R^4}+O(R^{-6}) \approx  f_{sss}+\frac{c^\prime_1}{N_s}+\frac{c^\prime_2}{N_s^2}+O(N_s^{-3}).
\end{aligned}
\end{equation}

The OPE coefficients involving spinning operator are slightly more complicated, since one has to carefully deal with the $\boldsymbol{\Omega}$ dependence. We compute $\int d \vec{\Omega} \bar{Y}_{2, 0}(\vec{\Omega})\left\langle\phi^{(1,0)}_{(0,0)}\left|[\vec{n}_\tau]^{(1,0)}(\vec{\Omega})\right|T^{(0,0)}_{(2,0)}\right\rangle=\left\langle\phi^{(1,0)}_{(0,0)}\left|[\vec{n}_\tau]^{(1,0)}_{(2,0)}\right| T^{(0,0)}_{(2,0)}\right\rangle$. The finite-size scaling of the OPE coefficient $f_{\phi\phi T}$ is given by
\begin{equation}
\begin{aligned}
 \sqrt{\frac{15}{8}} \frac{\left\langle\phi^{(1,0)}_{(0,0)}\left|[\vec{n}_\tau]^{(1,0)}_{(2,0)}\right| T^{(0,0)}_{(2,0)}\right\rangle}{\left\langle\phi^{(1,0)}_{(0,0)}\left|[\vec{n}_\tau]^{(1,0)}_{(0,0)}\right| 0\right\rangle} & \approx \frac{c_\phi f_{\phi \phi T} R^{-\Delta_\phi}+c_{\square \phi} f_{\phi, \square \phi, T} R^{-\left(\Delta_\phi+2\right)}+c_{\square^2 \phi} f_{\phi, \square^2 \phi, T} R^{-\left(\Delta_\phi+4\right)}+\cdots}{c_\phi R^{-\Delta_\phi}+c_{\square \phi} R^{-\left(\Delta_\phi+2\right)}+c_{\square^2 \phi} R^{-\left(\Delta_\phi+4\right)}+\cdots} \\
& \approx f_{\phi \phi T}+\frac{c_1}{R^2}+\frac{c_2}{R^{4}}+O(R^{-6})\approx f_{\phi \phi T}+\frac{c_1^{\prime}}{{N_s}}+\frac{c_2^{\prime}}{N_s^{2}}+O(N_s^{-3}).
\end{aligned}
\end{equation}

Similarly, $f_{\phi\phi J}$ and $f_{ss T}$can be computed by 
\begin{equation}
\begin{aligned}
\frac{\sqrt{3}\left\langle\phi^{(1,1)}_{(0,0)}\left|[\vec{n}_\tau]^{(1,0)}_{(1,0)}\right| J^{(1,1)}_{(1,0)}\right\rangle}{\left\langle\phi^{(1,0)}_{(0,0)}\left|[\vec{n}_\tau]^{(1,0)}_{(0,0)}\right| 0\right\rangle} & \approx \frac{c_\phi f_{\phi \phi J} R^{-\Delta_\phi}+c_{\square \phi} f_{\phi, \square \phi, J} R^{-\left(\Delta_\phi+2\right)}+c_{\square^2 \phi} f_{\phi, \square^2 \phi, J} R^{-\left(\Delta_\phi+4\right)}+\cdots}{c_\phi R^{-\Delta_\phi}+c_{\square \phi} R^{-\left(\Delta_\phi+2\right)}+c_{\square^2 \phi} R^{-\left(\Delta_\phi+4\right)}+\cdots} \\
& \approx f_{\phi \phi J}+\frac{c_1}{R^2}+\frac{c_2}{R^{4}}+O(R^{-6})\approx f_{\phi \phi J}+\frac{c_1^{\prime}}{{N_s}}+\frac{c_2^{\prime}}{N_s^{2}}+O(N_s^{-3}),
\end{aligned}
\end{equation}
and 
\begin{equation}
\begin{aligned}
 \sqrt{\frac{15}{8}} \frac{\left\langle s^{(0,0)}_{(0,0)}\left|[\vec{n}_\tau^2]^{(0,0)}_{(2,0)}\right| T^{(0,0)}_{(2,0)}\right\rangle}{\left\langle s^{(0,0)}_{(0,0)}\left|[\vec{n}_\tau^2]^{(0,0)}_{(0,0)}\right| 0\right\rangle} & \approx \frac{c_s f_{ss T} R^{-\Delta_s}+c_{\square s} f_{s, \square s, T} R^{-\left(\Delta_s+2\right)}+c_{\square^2 s} f_{s, \square^2 s, T} R^{-\left(\Delta_s+4\right)}+\cdots}{c_s R^{-\Delta_s}+c_{\square s} R^{-\left(\Delta_s+2\right)}+c_{\square^2 s} R^{-\left(\Delta_s+4\right)}+\cdots} \\
& \approx f_{ss T}+\frac{c_1}{R^2}+\frac{c_2}{R^{4}}+O(R^{-6})\approx f_{ss T}+\frac{c_1^{\prime}}{{N_s}}+\frac{c_2^{\prime}}{N_s^{2}}+O(N_s^{-3}),
\end{aligned}
\end{equation}
respectively.

\section{D. Two-point correlator}
In this section, we would like to study the correlator on the fuzzy sphere. We will use the operator $\phi$ as an example. 
The decomposition of local operator $\vec{n}_\tau$ is following Eq. \ref{eq:phi_decomposition}.
The normalized two-point function of $\vec{n}_\tau$ receives its leading order contribution from the two-point function of $\phi$: 
\begin{equation}
\begin{aligned}
G_{\phi\phi}\left(r,\theta\right) = & \frac{\left\langle 0\left|\vec{n}_\tau\left(r, \theta\right) \vec{n}_\tau\right| 0\right\rangle}{\left\langle \phi\left| \vec{n}_\tau\right|0\right\rangle^2} +O\left(R^{-1}\right)\\
= & \frac{\sum_{l=0}^{2 s} \bar{Y}_{l, 0}(\theta, 0) Y_{l, 0}(0,0)\left\langle0\left|[\vec{n}_\tau(r)]^{(1,0)}_{(l,0)} [\vec{n}_\tau]^{(1,0)}_{(l,0)}\right| 0\right\rangle}{\left\langle\phi\left|[\vec{n}_\tau]^{(1,0)}_{(0,0)}\right|0\right\rangle^2/(4\pi)}+O\left(R^{-1}\right), \\
= & \frac{r^{\Delta_\phi}}{\left(r^2+1-2 r \cos \theta\right)^{\Delta_\phi}}+O\left(R^{-1}\right).
\end{aligned}
\end{equation}
Fig. \ref{fig:Phi_2pt} depicts the two-point correlator $G_{\phi\phi}(r=1,\theta)$ by setting $r=1$ as a function of $\theta$. In this case, $G_{\phi\phi}(r=1,\theta)$ is a dimensionless function that solely depends on the angle $\theta$ between the two operator. Overall, the finite-size results approach theoretical expectation as $N_s$ increases. The discrepancy is relatively large at small angle. For $\theta \approx \pi/2$ (close to equator), the different curves almost merge together.


\begin{figure}[!htb]
	\includegraphics[width=0.35\textwidth]{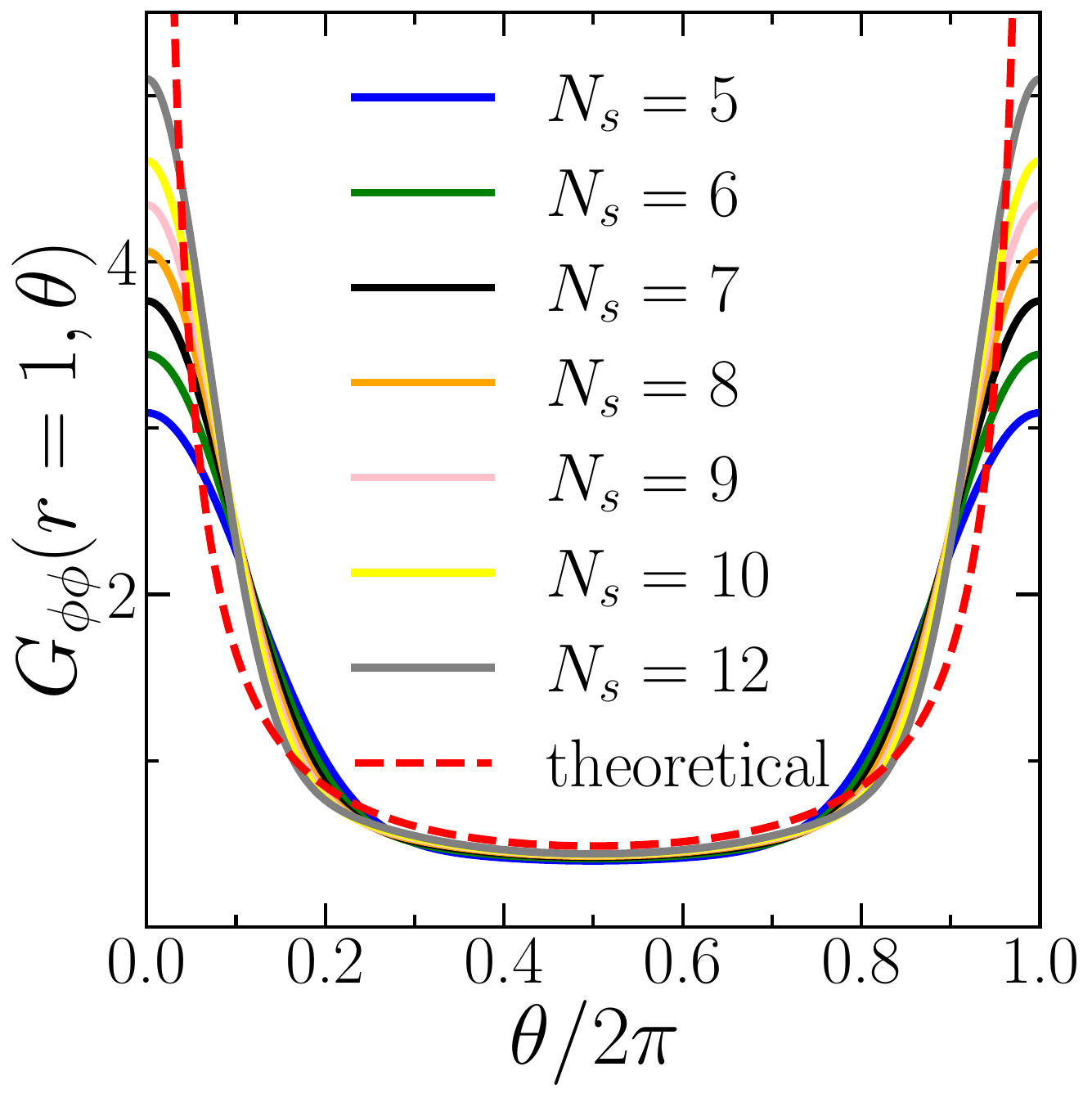}
	\caption{The angle dependence of the two-point correlator $G_{\phi\phi}(r = 1, \theta)$ is plotted for system sizes $N_s=5-12$.}
	\label{fig:Phi_2pt}
\end{figure}

\section{E. Details of numerical data}
\subsection{E.1 Raw numerical data}

\begin{table}
	\caption{Scaling dimensions of fields in $O(3)$ model ($\Delta \le 3$). The data from the fuzzy sphere model (FS) are rescaled by setting $\Delta_T=3$. Here $N_s=8$ data are from the ED and $N_s=9$ data are from the DMRG.	}
	\begin{tabular}{c|c|c|c|c|c}
		\hline
		\hline
		spin ($\ell$)  & charge ($S$) &Operator & Dimension (CB) & Dimension (FS) ($N_s=8$)  & Dimension (FS) ($N_s=9$) \\
		\hline		
		0 & 1 & $\phi$ & 0.519 & 0.521 & 0.522 \\	
		0 & 2 & $t_2$ & 1.210 &  1.244 & 1.243 \\	
		1 & 1 & $\partial_{\mu} \phi$ & 1.519 & 1.496 & 1.498 \\					
		0 & 0 & $s$ & 1.595 &  1.595 & 1.597 \\
		1 & 1 & $J_\mu$ & 2 & 2.038 & 2.032 \\
		0 & 3 & $t_3$ & 2.038 & 2.150 & 2.145 \\
		1 & 2 & $\partial_{\mu} t_2$ & 2.210 & 2.256 & 2.258 \\
		2 & 1 & $\partial_\mu \partial_{\nu} \phi$ & 2.519 & 2.504 & 2.529 \\
		0 & 1 & $\square \phi$ & 2.519 & 2.609 & 2.604 \\		
		1 & 0 & $\partial_\mu s$ & 2.595 & 2.643 & - \\
		0 & 4 & $t_4$ & 2.99 & 3.229 & 3.216 \\		
		2 & 0 & $T_{\mu\nu}$ & 3 & 3 & 3 \\
		2 & 1 & $\partial_{\nu} J_\mu$ & 3 & 3.138 & - \\
		1 & 1 & $\epsilon_{\mu \nu \rho} \partial_\nu J_\rho$ & 3 & 3.018 & - \\
		\hline
		\hline 
	\end{tabular}\label{tab:listraw}
\end{table}

In this section, we present the data of energy spectra and corresponding conformal multiplet of various fields in Tab. \ref{tab:listraw}. Here we only present the data for relevant fields ($\Delta\le 3$). 
Importantly, we emphasize that, the operator spectrum doesnot miss any CFT field or contain any extra non-CFT state in the regime ($\Delta \le 3$). 
These data are used for plotting the figures in the main text. 
For comparison, we also list the results from conformal boostrap method \cite{Su2021,RMP_CB}.
These detailed data  gives a good quantification for the numerical error. That is, scaling dimensions of the low-lying fields are quite close to the results of conformal bootstrap, and the trend towards thermodynamic limit is correct for most of fields. 
A rigorous error analysis based on the finite size scaling will be presented in the next subsection.

Another interesting point is that, almost perfect state-operator
correspondence is manifest in surprisingly small system sizes, e.g. the numerical data at a given system size $N_s=8$, which is the largest system size that we can reach using ED, is already close to the 3D CFT data. Here, to further elucidate that the numerical findings indeed reflect the physics in the thermodynamic limit, we show the energy spectra on different system sizes.   
As one can see that, the energies on different system sizes match the prediction of 3D CFT quite well.

A part of data can be accessed in the DMRG computation. 
In the DMRG calculation, we explicitly implement three U(1) symmetries, i.e. z-component angular momentum quantum number $L^z$, total electron number $n^\uparrow + n^\downarrow$, and z-component spin  $S^z=n^\uparrow - n^\downarrow$. 
Making use of the symmetry information of the different fields, we can simplify the DMRG calculation. For example, to target the lowest $\mathrm O$(3) vector field $\phi$, we can calculate the lowest energy state in the sector $L^z=0,S^z=1$ (instead of directly targeting higher excited states in $L^z=0, S^z=0$). Accessing the ground state in different symmetry sectors usually gets fast convergence compared with targeting the excited states.   
For the DMRG calculations, we only focus on the low-lying fields, so some of data are missing in Tab. \ref{tab:listraw} (last column).


\subsection{E.2  Error analysis} 

At last, we present an error analysis of  obtained scaling dimensions. Generally, the typical length scale $R$, the radius of fuzzy sphere, should be scaled with the number of Landau orbitals (i.e. spins) as $R\sim \sqrt{N_s}$. $R$ behaves as the typical length scale $L_x$ as in the flat spacetime lattice model. To extrapolate the numerical data to the thermodynamic limit $R\rightarrow \infty$, we use the polynomial function
$f_{O}(R) = f_O(\infty) + \frac{a_1(O)}{R} + \frac{a_2(O)}{R^2} + O(\frac{1}{R^2}) $, and fit the finite-size data using least-square method. The mean values $f_O(\infty)$ give the best estimate of the scaling dimensions in the thermodynamic limit, and the residual give the relative errors (See Tab. \ref{tab:ErrorDim}).

\begin{table}[!h]
	\setlength{\tabcolsep}{0.2cm}
	\renewcommand{\arraystretch}{1.4}
	\centering
	\caption{Extrapolated scaling dimensions of low-lying primary operators identified via state-operator correspondence on the fuzzy sphere. \label{tab:ErrorDim}}
	\begin{tabular}{cccccc} \hline\hline
		& $\phi$ & $t_2$  & $s$ & $t_3$ & $t_4$ \\
		\hline
		Fuzzy sphere & 0.524$\pm$0.004 & 1.211$\pm$0.008 & 1.588$\pm$0.009 & 2.028$\pm$0.011 & 2.961$\pm$0.012 \\			
		\hline\hline
		
	\end{tabular} 
\end{table}

Next, we further analyze the error of obtained OPE coefficients. The strategy used to estimate errors are explained below. Following the discussion in Ref. \cite{Hu2023}, we utilize the different local operators to estimate the OPE coefficients. (The general idea is \cite{Zou2020}, the information of CFT primary field may be encoded in different local operators, and it is expected different local operators should give the same OPE coefficients since the OPE coefficients should be universal.)    
First of all, we need to analyze the operator content of different local operators based on the symmetries. In short, the following local operators have significant weights with the primary fields $s, \phi$:
\begin{align}
&\mathrm{O(3) vector}: \phi \sim \vec{n}_\tau(\vec{\Omega}) \\
&\mathrm{O(3) vector}: \phi \sim \vec{n}^2_\tau(\vec{\Omega}) \vec{n}_\tau(\vec{\Omega}) \\
&\mathrm{O(3) vector}: \phi \sim O^s(\vec{\Omega}) \vec{n}_\tau(\vec{\Omega}) \\
&\mathrm{O(3) scalar}: s\sim \vec{n}^2_\tau(\vec{\Omega}) \\
&\mathrm{O(3) scalar}: s\sim O^s(\vec{\Omega}) = n(\vec{\Omega})n(\vec{\Omega})+0.015*n^{\tau^x\otimes \sigma^0}(\vec{\Omega}) 
\end{align}

Using the local operators listed above, we can estimate the OPE coefficients. For example, when estimating the OPE coefficient $f_{\phi\phi s}$, one could use two different ways to calculate it: 1) $\langle \phi |\phi | s\rangle $ (see Eq. \ref{eq:phiphis}) by using the local operator $\phi \sim \vec{n}_\tau(\vec{\Omega})$, $\phi \sim \vec{n}^2_\tau(\vec{\Omega}) \vec{n}_\tau(\vec{\Omega})$, or $O^s(\vec{\Omega}) \vec{n}_\tau(\vec{\Omega}) $ and 2)  $\langle \phi |s|\phi \rangle $ (see Eq. \ref{eq:phisphi}) by using the local operator $s\sim \vec{n}^2_\tau(\vec{\Omega})$ or $s\sim O^s(\vec{\Omega}) $. 
Fig. \ref{fig:OPE_Error1} shows the five different estimations. By extrapolation, the estimated values of $f_{\phi\phi s}$ are 0.518 from $\langle \phi |\vec{n}_{\tau}| s\rangle$, 0.529 from $\langle \phi |\vec{n}^2_\tau|\phi \rangle$, 0.532 from $\langle \phi |O^s|\phi \rangle$, 0.530 from $\langle \phi |\vec{n}^2_\tau\vec{n}_\tau|\phi \rangle$, and 0.515 from $\langle \phi |O^s\vec{n}_\tau|\phi \rangle$. So the mean value and relative error are estimated to be
\begin{align}
 f_{\phi\phi s} \approx 0.525 \pm 0.007.
\end{align}
Similarly, the estimated values of $f_{sss}$ are 0.498 from $\langle s|\vec{n}^2_\tau|s\rangle$ and 0.517 from $\langle s|O^s|s\rangle$  (see Fig. \ref{fig:OPE_Error1}). So the mean value and relative error are given by 
\begin{align}
f_{sss} \approx 0.507 \pm 0.010.
\end{align}
Meanwhile, the estimated values of $f_{ssT}$ are 0.594 from $\langle s|\vec{n}^2_\tau|T\rangle$ and 0.563 from $\langle s|O^s|T\rangle$ (see Fig. \ref{fig:OPE_Error1}). Their mean value and relative error are given by 
\begin{align}
f_{ssT} \approx 0.578 \pm 0.016.
\end{align}

The estimated values of $f_{\phi\phi J}$ are 0.768 from $\langle \phi|\vec{n}_\tau|J\rangle$, 0.772 from $\langle \phi|\vec{n}^2_\tau\vec{n}_\tau|J\rangle$, and 0.715 from $\langle \phi|O^s\vec{n}_\tau|J\rangle$ (see Fig. \ref{fig:OPE_Error1}). Their mean value and relative error are given by 
\begin{align}
f_{\phi\phi J} \approx 0.752 \pm 0.025.
\end{align}

The estimated values of $f_{\phi\phi T}$ are 0.164 from $\langle \phi|\vec{n}_\tau|T\rangle$, 0.164 $\langle \phi|\vec{n}^2_\tau\vec{n}_\tau|T\rangle$, and 0.163 from $\langle \phi|O^s\vec{n}_\tau|T\rangle$ (see Fig. \ref{fig:OPE_Error1}). Their mean value and relative error are given by 
\begin{align}
f_{\phi\phi T} \approx 0.1685 \pm 0.0003.
\end{align}
We find the estimates of $f_{\phi\phi T}$ from three different local operators are quite close to each other, so the relative error is much smaller. 

At last, we need to emphasize, although the above errors are not rigorous, we think this is the best way to do the error analysis \cite{Hu2023}. That is, compared with the error in the fitting process, the relative errors from the fitting using different local operators are relative larger. So we would like to use this way to estimate the error of OPE coefficients.

\begin{table}[!h]
	\setlength{\tabcolsep}{0.2cm}
	\renewcommand{\arraystretch}{1.4}
	\centering
	\caption{Extrapolated OPE coefficients of low-lying primary operators identified via state-operator correspondence on the fuzzy sphere. }
	\begin{tabular}{cccccc} \hline\hline
		& $f_{\phi\phi s} $ & $f_{sss}$  & $f_{ssT}$ & $f_{\phi\phi J}$ & $f_{\phi\phi T}$ \\
		\hline
		Fuzzy sphere & 0.525$\pm$0.007 & 0.507$\pm$0.010 & 0.578$\pm$0.016 & 0.752$\pm$0.025 & 0.1685$\pm$0.0003 \\			
		\hline\hline
		
	\end{tabular} 
\end{table}

\begin{figure}[!htb]
	\includegraphics[width=0.3\textwidth]{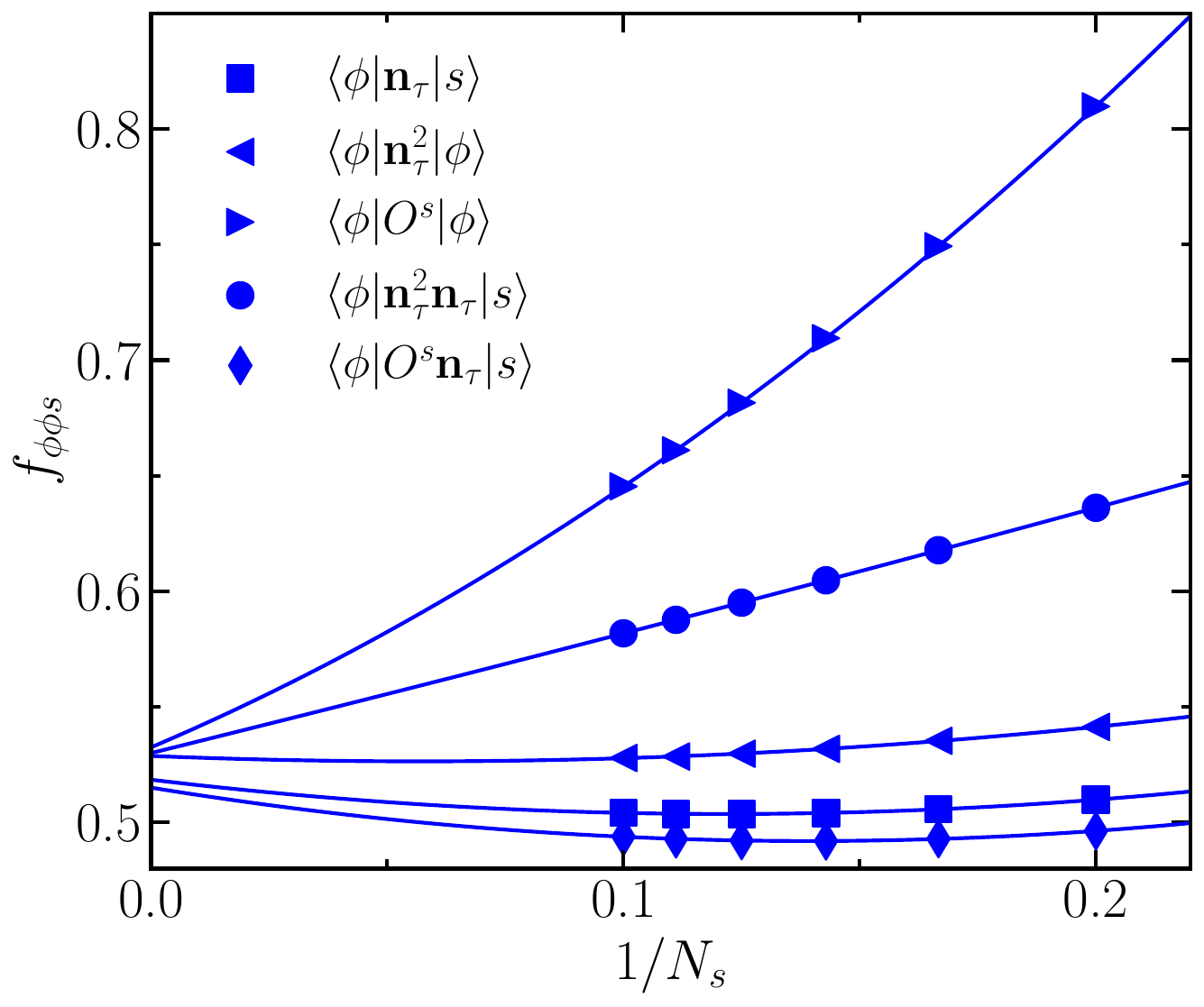}
    \includegraphics[width=0.3\textwidth]{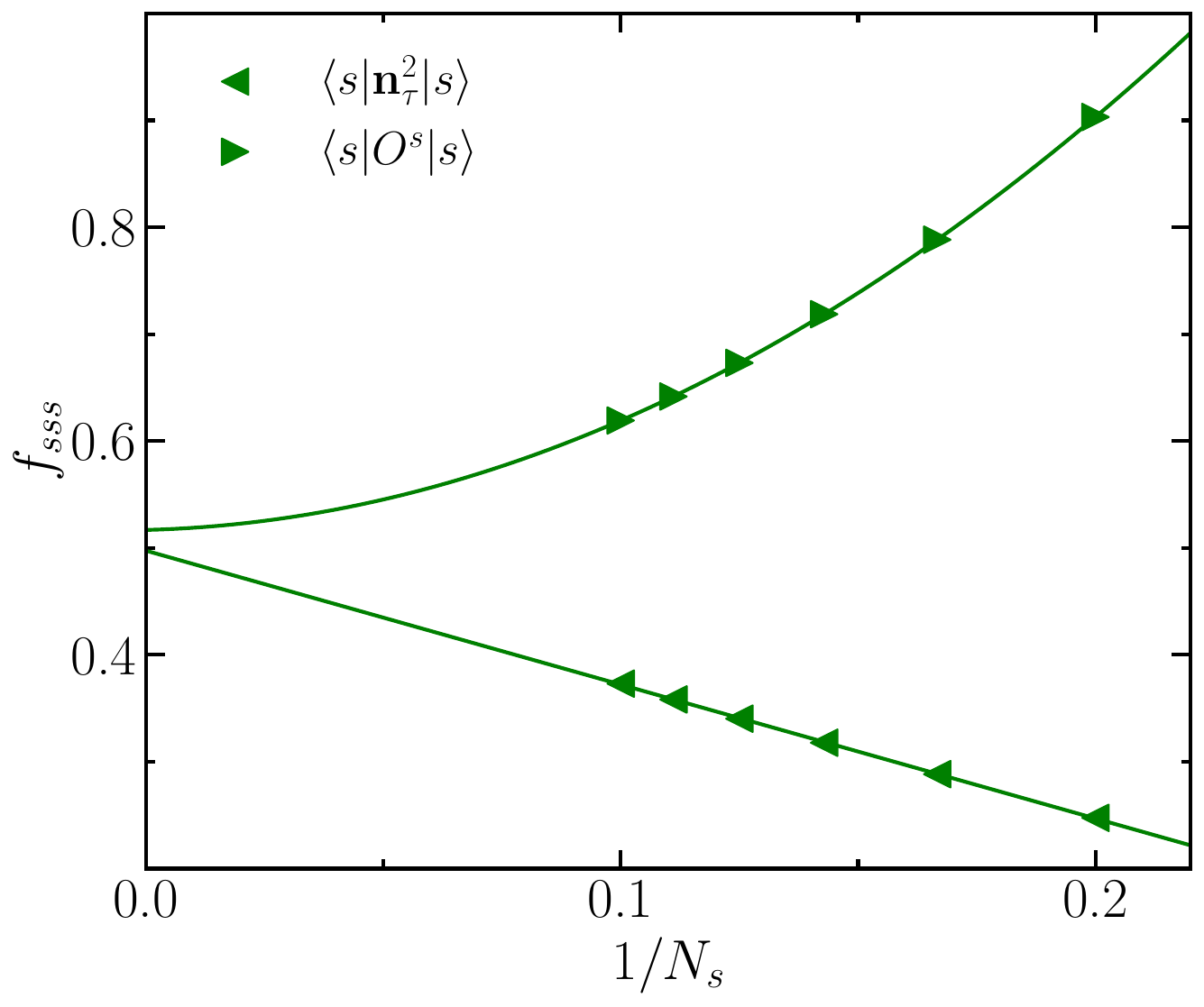}
    \includegraphics[width=0.3\textwidth]{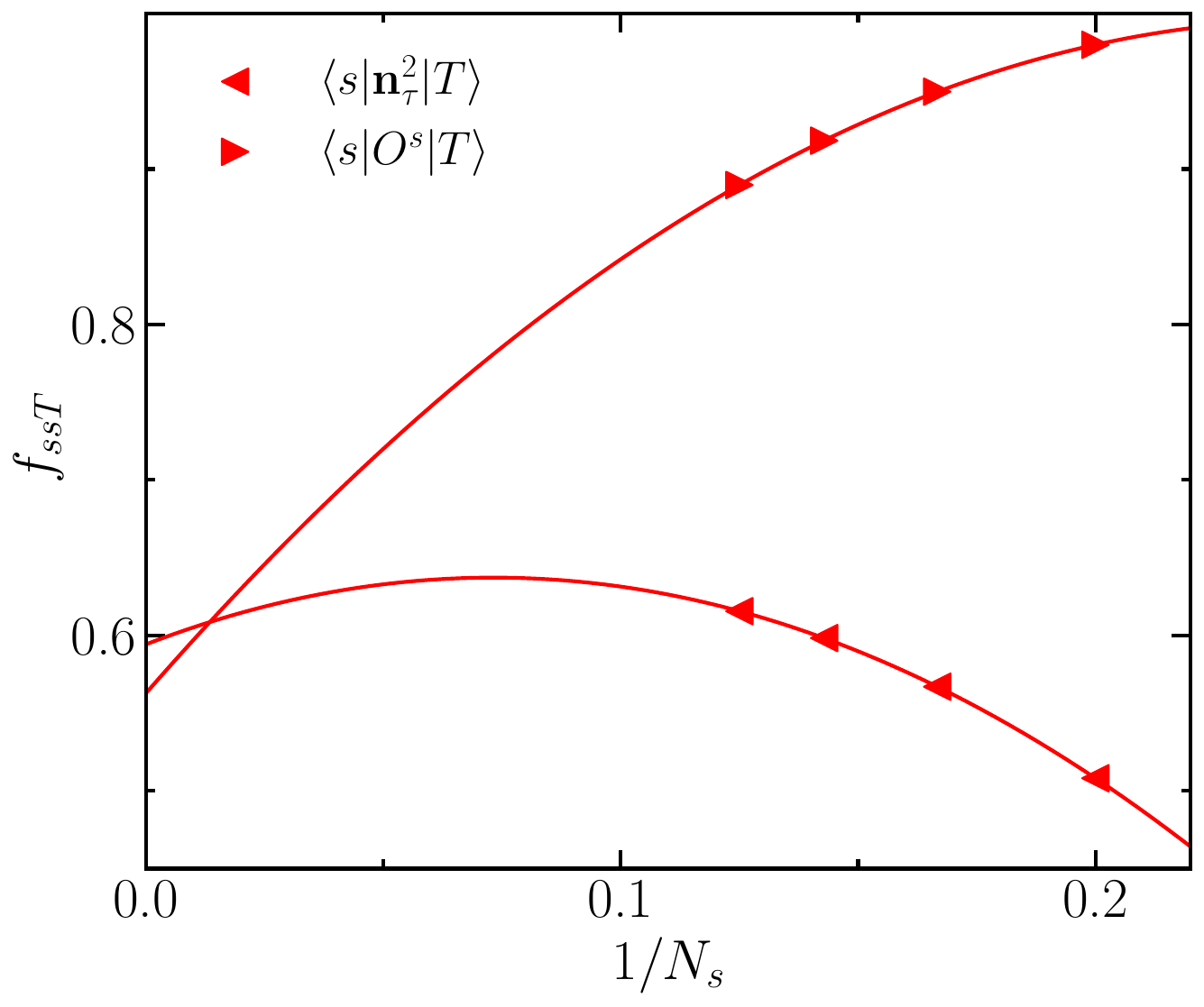}
    \includegraphics[width=0.3\textwidth]{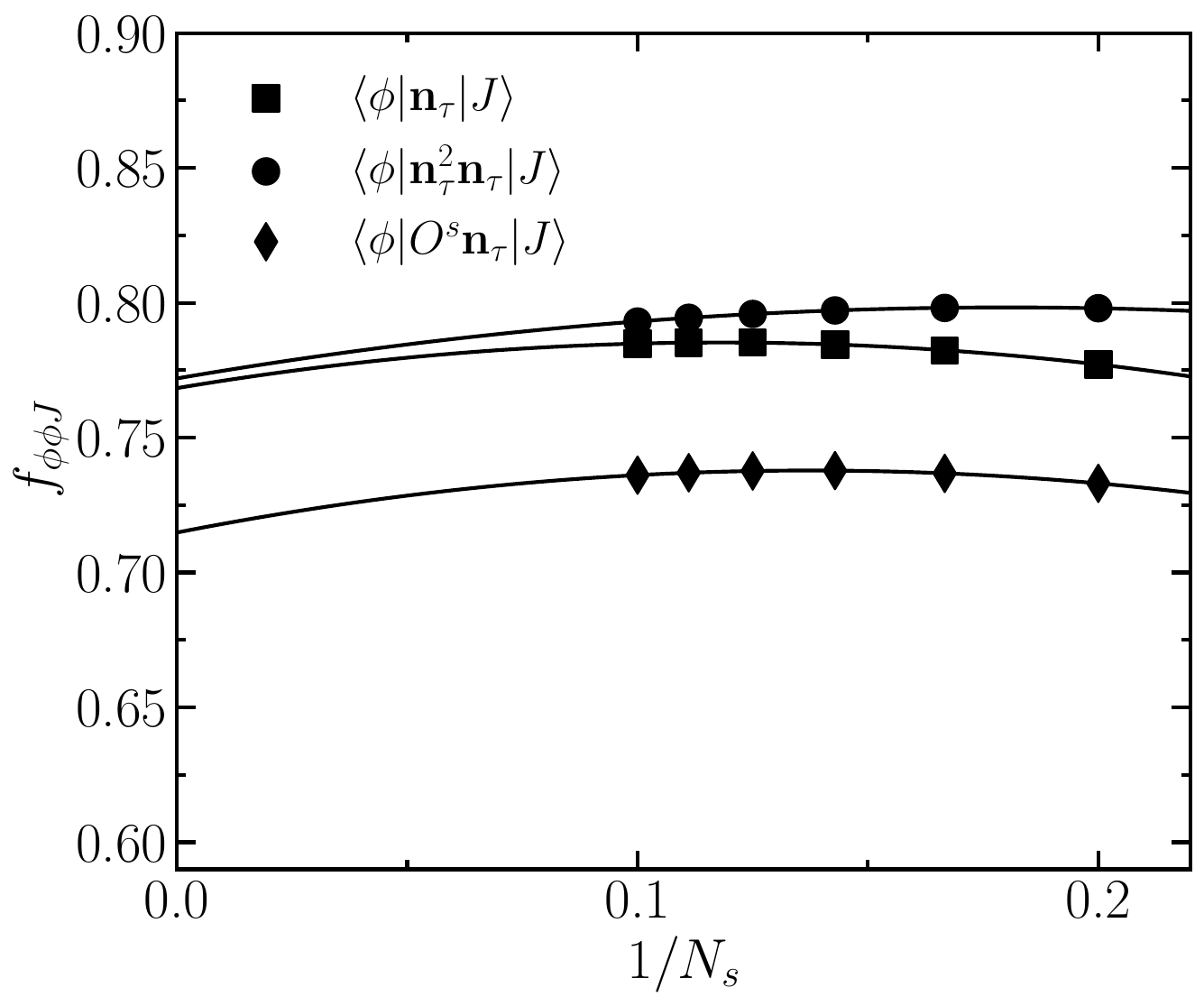}
    \includegraphics[width=0.3\textwidth]{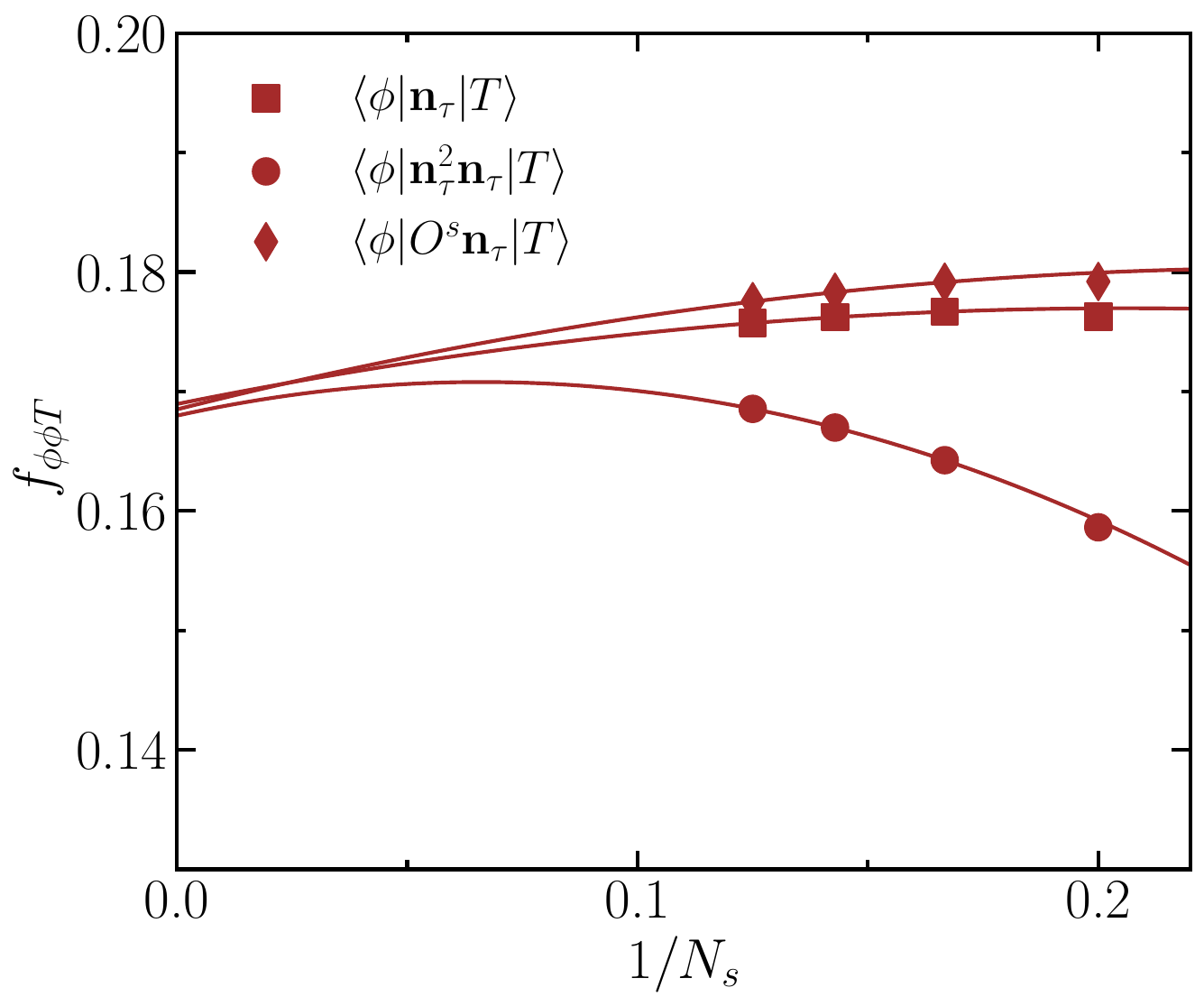}
	\caption{(Top left)
 Finite-size scaling of five different quantities to estimate the OPE coefficient $f_{\phi\phi s}$.
 (Top middle) 
 The OPE coefficient $f_{sss}$ can be estimated by two different local operators $\vec{n}^2(\Omega)$ and $O^s$. 
 (Top right) The OPE coefficient $f_{ssT}$ can be estimated by two different local operators $\vec{n}^2(\Omega)$ and $O^s$. 
 (Bottom left)
 Finite-size scaling of three different quantities to estimate the OPE coefficient $f_{\phi\phi J}$.
 (Bottom right) The OPE coefficient $f_{\phi\phi T}$ can be estimated by two different local operators $\vec{n}^2(\Omega)$ and $O^s$.
 }
	\label{fig:OPE_Error1}
\end{figure}

\end{widetext}

\end{document}